\documentclass[useAMS,usenatbib]{mn2e}

\usepackage{epsfig,graphicx,mathptm}
\usepackage{times}
\usepackage{natbib}

\newcommand{\gtrsim}{\ga}
\newcommand{\lesssim}{\la} 

\newcommand{\sbunits}{erg s$^{-1}$ cm$^{-2}$ deg$^{-2}$}

\newcommand{\ypar}{$y$-parameter}
\newcommand{\bpar}{$b$-parameter}
\newcommand{\lcdm}{$\Lambda$CDM}
\newcommand{\sph}{\textsc{sph}}

\def\prd{Phys. Rev. D}
\def\aap{A\&A}
\def\apj{ApJ}

\def\apjl{ApJ}
\def\mnras{MNRAS}
\def\araa{ARA\&A}

\def\physrep{Phys. Rep.}

\def\apjs{ApJS}

\title[The SZ effects from a cosmological simulation] 
{The Sunyaev-Zel'dovich effects from a cosmological
hydrodynamical simulation: large-scale properties and correlation
with the soft X-ray signal}
\author[M. Roncarelli et al.]
{M. Roncarelli$^1$,
L. Moscardini$^{1,2,3}$, 
S. Borgani$^{4,2,3}$ and 
K. Dolag$^5$ \\
$^1$ Dipartimento di Astronomia, Universit\`a di Bologna, via Ranzani
  1, I-40127 Bologna, Italy (mauro.roncarelli, lauro.moscardini@unibo.it)\\
$^2$ INAF -- National Institute for Astrophysics, Italy \\
$^3$ INFN -- National Institute for Nuclear Physics, Italy \\
$^4$ Dipartimento di Astronomia, Universit\`a di Trieste, via
  Tiepolo 11, I-34131 Trieste, Italy (borgani@oats.inaf.it) \\
$^5$ Max-Planck-Institut f\"ur Astrophysik, Karl-Schwarzschild Strasse
  1, D-85741 Garching bei M\"unchen, Germany (kdolag@mpa-garching.mpg.de)
}

\begin{document}


\pagerange{\pageref{firstpage}--\pageref{lastpage}} \pubyear{2007}

\maketitle

\label{firstpage}

\begin{abstract}
Using the results of a cosmological hydrodynamical simulation of the 
concordance \lcdm\ model, we study the global properties of the 
Sunyaev-Zel'dovich (SZ) effects, both considering the thermal (tSZ) and 
the kinetic (kSZ) component. The simulation follows gravitation and gas 
dynamics and includes also several physical processes that 
affect the baryonic component, like a simple reionization 
scenario, radiative cooling, star formation and supernova feedback. Starting from 
the outputs of the simulation we create mock maps of the SZ signals due 
to the large structures of the Universe integrated in the range $0 \leq z 
\leq 6$. We predict that the Compton \ypar\ has an average value of 
$(1.19 \pm 0.32) \times 10^{-6}$ and is lognormally distributed in the sky; 
half of the whole signal comes from $z<1$ and about 10 per cent from $z>2$. 
The Doppler \bpar\ shows approximately a normal distribution with vanishing 
mean value and a standard deviation of $1.6 \times 10^{-6}$, with a 
significant contribution from high-redshift ($z>3$) gas. We find 
that the tSZ effect is expected to dominate the primary CMB anisotropies for 
$\ell \gtrsim 3000$ in the Rayleigh-Jeans limit, while interestingly the 
kSZ effect dominates at all frequencies at very high multipoles 
($\ell \gtrsim 7 \times 10^4$). 
We also analyse the cross-correlation between the two SZ 
effects and the soft (0.5--2 keV) X-ray emission from the intergalactic 
medium and we obtain a strong correlation between the three signals, 
especially between X-ray emission and tSZ effect ($r_\ell \simeq $ 
0.8-0.9) at all angular scales.
\end{abstract}



\section{Introduction} \label{sect:intro}

In the recent years the cosmic microwave background (CMB) primordial
fluctuations have been measured with high accuracy thanks to the
observations made by the {\it Wilkinson Microwave Anisotropy Probe}
satellite \citep[{\it WMAP},][]{spergel2003,spergel2006}, establishing
an important step forward in precision cosmology. The primary CMB
anisotropies, in fact, constitute the most important dataset to obtain
information about the physical conditions of the early Universe and,
together with other results coming from high-redshift supernovae
\citep[see, e.g.,][]{astier2006,wood-vasey2007}, weak lensing 
\citep[see, e.g.,][]{heymans2005,massey2005,hetterscheidt2006,semboloni2006,hoekstra2006} 
and galaxy clustering
\citep[see, e.g.,][]{cole2005,eisenstein2005,tegmark2006,sanchez2006}, they
indicated as a favorite scenario a flat cosmological model dominated
by a cosmological constant: the so-called \lcdm\ model.

The dynamical and thermodynamical effects of the large scale structure
(LSS) formation after the recombination also create secondary
anisotropies in the CMB signal. Most importantly, after the epoch of
reionization the CMB photons interact with the free electrons of the
intergalactic medium (IGM) via Thomson scattering giving rise to the
Sunyaev-Zel'dovich (SZ) effect \citep{sunyaev1972,sunyaev1980}: this
constitutes both a noise for the primary CMB signal and a probe for
the baryon physics. The SZ effect is usually classified into two
components: the dominant one, the thermal SZ (tSZ) effect, is the
inverse-Compton scattering caused by the thermal motion of a
population of high-temperature electrons, mainly located in the hot
plasma of galaxy clusters, which results in a gain in energy for the
photons and a consequent distortion of the black-body spectrum of the
CMB. Conversely, the kinetic SZ (kSZ) effect is the Doppler shift
caused by the bulk motion of the ionized gas and can result either in
a gain or in a loss of photon energy, depending on the direction of
the gas velocity with respect to the photon.

After its first claimed detection by \cite{parijskij1973} and the 
works of \cite{birkinshaw1991} and \cite{birkinshaw1994}, the
development of new microwave instruments has now made available a
large amount of observational data for the study of the tSZ effect in
galaxy clusters, including two-dimensional mapping. In the near future
this observational field is believed to receive a significant boost
thanks to a new generation of suitable land-based instruments, like
{\it AMI}, {\it SPT}, {\it ACT}, {\it SZA}, {\it AMiBA} and 
{\it APEX}, and the launch of the {\it Planck}
satellite that will provide a full-sky catalogue of galaxy clusters
detected via the tSZ effect.  Finally, the project of the {\it Atacama 
Large Millimeter Array} ({\it ALMA}) specifically includes 
high-resolution imaging of the SZ effect in its scientific goals.

The development of the SZ science for galaxy clusters will be strictly
connected with the already available data in the X-ray band that
nowadays constitute the most important dataset for cluster physics. In
fact, since both the bremsstrahlung emission and the SZ effect
depend on the density and temperature of the gas, the two
signals are expected to be highly correlated and their comparison will
be of great interest in order to understand the systematics of the two
observables.  In addition to it, while the X-ray signal has proven to
be a sensitive probe for the properties of the hot gas in the central
regions of nearby clusters, the absence of redshift dimming for the SZ
effect and its different dependence on the density will allow one to
detect more distant sources and, hopefully, to obtain more information
on the external regions of galaxy clusters where the modelization of
the cluster physics has smaller uncertainties \citep[see,
e.g.,][]{roncarelli2006b}. Besides from cluster physics the tSZ effect
arising from the whole LSS, which is expected to be the most important
signal at the angular scales of some arcminutes, can also be used to
directly constrain cosmology because of the strong dependence of its 
amplitude on the normalization of the power spectrum $\sigma_8$ 
\citep[see, e.g.,][]{goldstein2003,bond2005,douspis2006}. On the 
other side a measurement of the
kSZ effect can yield information on the peculiar velocity field at
high redshift and, consequently, put constraints on different dark
energy models \citep[see, e.g.,][]{hernandez2006} and on cosmological 
parameters in general \citep[see, e.g.,][]{bhattacharya2007}; at the 
second order
the kSZ effect is also believed to be marginally affected by the
dynamical effects associated with epoch of reionization, even if a
future measurement of this signal appears very challenging \citep[][]
{iliev2006}.

In this framework a theoretical analysis of the SZ effect and of its
correlation with the X-ray signal is crucial not only for
investigating the connections between these observables and the LSS
formation, but also for modeling and, possibly, taking under control
the SZ-noise, that is expected to affect high-multipole CMB
observations.  In this paper we present our results on the statistical
properties of the SZ signal obtained from a cosmological
hydrodynamical simulation \citep{borgani2004} that follows the
evolution of the structure formation in the framework of the
\lcdm\ cosmology. Since the thermodynamical evolution of the baryons 
depends not only on the cosmological parameters but also on different
non-thermal processes influencing the gas physics, our numerical model
specifically includes several physical effects connected with
radiative cooling and star formation that have a 
significant effect on the shape of the temperature and density
profiles of the inner regions of galaxy clusters and, consequently, on
their X-ray and SZ properties.

This paper is organized as follows. In the next Section we briefly 
review the basic equations that describe the tSZ and kSZ effects. In Section 
\ref{sect:models} we present the characteristics of our cosmological 
simulation (Section \ref{sect:simulation}) and the method used to
obtain predictions on the SZ effects from its outputs (Section
\ref{sect:cones}). In Section \ref{sect:signal} we discuss the average
large-scale properties of the SZ effects. In Section
\ref{sect:power_sp} we analyze the power spectra of the tSZ and kSZ
signals and in Section \ref{sect:cross} we study their
cross-correlation. Section \ref{sect:X-ray} is devoted to the
description of the correlation properties with the soft X-ray
signal. Finally we summarize our conclusions in Section
\ref{sect:conclu}.


\section{Basics of the SZ effects} \label{sect:sze}

Here we review some of the basic equations for both SZ effects in the
non-relativistic approximation.  More details can be found in
several reviews \citep[see,
e.g.,][]{rephaeli1995,birkinshaw1999,carlstrom2002,rephaeli2005}.

The intensity of the tSZ effect in a given direction 
is usually expressed in terms of the Compton 
\ypar\ defined as
\begin{equation}
y \equiv \frac{k_B\, \sigma_T}{m_e\, c^2} \int{dl\, n_e(T_e-T_{\rm CMB})} \ ,
\label{eq:ypar}
\end{equation}
where $m_e$ is the electron rest mass, $c$ is the light speed, $n_e$
and $T_e$ are the electron number density and temperature, and $T_{\rm
CMB}$=2.726 K is the CMB temperature \citep{mather1994}. The resulting
change in the latter due to the scattering of the electrons is
directly proportional to the value of the \ypar\ and is given by
\begin{equation}
\frac{\Delta T}{T_{\rm CMB}} = y \, g_\nu(x)\ ,
\label{eq:dt_t}
\end{equation}
where $x \equiv h\, \nu /(k_B T_{\rm CMB})$ is the dimensionless
frequency and $g_\nu(x)$ represents the dependence on the 
observation frequency:
\begin{equation}
g_\nu(x) = \left( x\, \frac{e^x+1}{e^x-1}-4 \right) \ .
\label{eq:g_nu}
\end{equation}
It is important to note that in the Rayleigh-Jeans (RJ) limit 
($x \ll 1$) this expression reduces to $g_\nu(x) \simeq -2$ and that for 
$\nu \simeq 218$ GHz the tSZ effect is null.

The kSZ effect can be expressed in terms of the Doppler \bpar\ defined
as
\begin{equation}
b \equiv \frac{\sigma_T}{c}\int{dl\, n_e\, v_r} \ ,
\label{eq:bpar}
\end{equation}
where $v_r$ is the radial component of the peculiar velocity of the
gas element (positive if it is moving away from the observer, negative
if it is approaching). The resulting measured temperature fluctuation
is $\Delta T/T_{\rm CMB} = -b $. Note that, unlike for the tSZ effect, 
this is independent of the observation frequency.


\section{Models and method} \label{sect:models}

It is known that the global properties of the tSZ and kSZ effects
depend on the physical characteristics of the LSS as a
whole. Moreover, as we will discuss in Section \ref{sect:signal},
significant contributions to both SZ effects comes from high-redshift
gas. Therefore a theoretical study of their global properties and the
comparison with present and future observations require a realistic 
modelization of the thermodynamical and dynamical history of the gas
filling the volume of the past light-cone seen by an observer located
at $z=0$ out to the epoch of reionization.  For this purpose we use the
outputs of a cosmological hydrodynamical simulation at different 
redshifts to build different realizations of light-cones, which
enable the production of simulated maps of the SZ signals.


\subsection{The cosmological hydrodynamical simulation}  \label{sect:simulation}

We use the results of the cosmological hydrodynamical simulation by 
\cite{borgani2004}, which considers the concordance
cosmological model, i.e. a flat \lcdm\ model dominated by the
presence of the cosmological constant ($\Omega_{\rm m}=0.3$,
$\Omega_{\Lambda}=0.7$), with a Hubble parameter $h\equiv H_0/$(100
km s$^{-1}$ Mpc$^{-1}$)=0.7, and a baryon density $\Omega_{\rm
b}=0.04$.  The initial conditions were generated by sampling from a
cold dark matter (CDM) power spectrum, normalized by assuming
$\sigma_8=0.8$, being $\sigma_8$ the r.m.s. matter fluctuation into a
sphere of radius $8 h^{-1}$ Mpc.  The run, that was carried out with
the \textsc{tree-sph} code \textsc{gadget-2}
\citep{springel2001,springel2005}, followed the evolution of $480^{3}$
dark matter (DM) particles and as many gas particles from $z=49$ to
$z=0$.  The side of the cubic box is $192 h^{-1}$ Mpc and
correspondingly the masses of the DM and gas particles are $m_{\rm
DM}= 4.62 \times 10^9 h^{-1} M_\odot$ and $m_{\rm gas}= 6.93 \times
10^8 h^{-1} M_\odot$, respectively.  The Plummer-equivalent
gravitational softening of the simulation was set to $\epsilon$=7.5
$h^{-1}$ kpc in physical units between $z=2$ and $z=0$, and fixed in
comoving units at higher redshifts. This run produced one hundred
outputs, equally spaced in the logarithm of the expansion factor,
between $z=9$ and $z=0$.

The simulation treats not only gravity and non-radiative
hydrodynamics, but also includes different processes that can 
influence the physics of the intracluster medium (ICM), like star 
formation \citep[by adopting a
sub-resolution multiphase model for the interstellar medium; see
][]{springel2003}, feedback from type II supernovae (SN-II) with 
the effect of galactic outflows, radiative cooling processes 
within an optically 
thin gas of hydrogen and helium in collisional ionization equilibrium 
and heating \citep[by a uniform, time-dependent, photoionizing UV 
background expected from a population of quasars and modelled 
following][]{haardt1996}.

The results of this simulation have been used for different previous
investigations \citep[see,
e.g.,][]{murante2004,ettori2004,cheng2005,rasia2005}. \cite{diaferio2005}
used the whole sample of galaxy clusters extracted from this
simulation at $z=0$ to find a good agreement with the observed scaling
relations between X-ray and SZ properties and to evaluate the lower
limit (about 200 km s$^{-1}$) to the systematic errors that will
affect future measurements of cluster peculiar velocities with the SZ
effect.  Recently \cite{roncarelli2006a} used
this simulation and the same light-cone realizations to estimate the
soft (0.5--2 keV) X-ray emission arising from the diffuse gas and found
a good agreement with the current upper limits of the unresolved X-ray
background.


\subsection{The map-making procedure}  \label{sect:cones}

In order to create mock maps of the tSZ and kSZ signals produced by
the LSS integrated over redshift, we use the same
light-cone simulations analyzed by \cite{roncarelli2006a} for the 
study of the X-ray emission. Note that similar applications of the 
same technique, also in terms of SZ effects, have been done by 
different authors \citep[see, e.g.,][]{dasilva2000,croft2001,
dasilva2001a,dasilva2001b,springel2001,white2002,zhang2002}. The 
method is based on the replication of the original box volume along 
the line of sight. For the projection we prefer to make use of the 
comoving coordinates, taking advantage of the fact that in a flat 
cosmology, like the one here assumed, light travels along a straight 
line. We build past light-cones extending out to $z=6$. As we will 
discuss in more detail in Section \ref{sect:signal}, this limit is 
enough to account for almost all of the tSZ signal; on the contrary 
the kSZ effect is believed to have a non-negligible contribution from the 
gas located at $z>6$, out to the epoch of reionization. Anyway, since 
the reionization model assumed in our simulation considers the gas at 
$z\gtrsim 6$ almost neutral, extending the calculation to higher 
redshifts would not produce a significant change in the kSZ signal. 
Therefore we stress that the results for the kSZ effect presented in 
this work \emph{cannot} be considered as the whole signal expected, 
but rather as the signal arising from the gas in the range $0 \leq z 
\leq 6$ which, according to the results of 
\cite{iliev2006}, should constitute the dominating component. The 
extension of our light-cone realizations corresponds to a comoving 
distance of approximately $5770\ h^{-1}$ Mpc, so we would need to stack 
the simulation volume roughly 30 times.  However, in order to obtain a
better redshift sampling, rather than stacking individual boxes, we
adopt the following procedure.  Each simulation volume 
necessary to construct the light-cone is divided along the line of 
sight into three equal slices,
each of them having a depth of $64\ h^{-1}$ Mpc. Then the stacking
procedure is done by choosing the slice extracted from the simulation
output that best matches the redshift of the central point of the
slice. Our light-cones are thus built with 91 slices extracted from 82
different snapshots.

\begin{figure*}
\includegraphics[width=0.45\textwidth,angle=270]{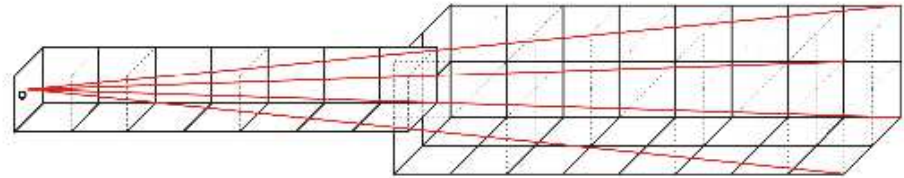}
\caption{
Sketch of the configuration adopted to realize the light-cones.  The
observer is located at the position $O$ at the centre of the left
side of the first box.  The past light-cone is obtained by stacking the
comoving volumes of the simulation outputs taken at the corresponding
redshift.  In order to obtain a large field of view of size $3.78^2$
deg$^2$, starting at $z\simeq1.4$ we use four replications of the
box at the same redshift.  The red lines show the volume inside the 
light cone corresponding to the field of view.  }
\label{fig:cubes}
\end{figure*}

In order to avoid that the same structures are repeated along the line
of sight, we include a process of box randomization in our map-making
procedure, taking advantage of the periodic boundary conditions of our
simulation: for each box entering the light-cone we combine a process
of random recentring of the coordinates with a 50 per cent
probability of reflecting each axis.  In order to avoid spatial
discontinuities between slices belonging to the same box, we impose
the same randomization process to them\footnote{Adopting this 
procedure, 
the surfaces separating slices belonging to the same box do not 
correspond to true spatial discontinuities but rather to redshift 
discontinuities.}: in this way we can retain the
global information on the large-scale structures present in the box
and at the same time we strongly reduce the loss of large-scale power.
In order to extend the field of view of our maps, we replicate the
boxes four times across the line of sight starting at comoving
distances larger than half the light-cone extension (i.e. larger than
about $2900\ h^{-1}$ Mpc, corresponding to $z\gtrsim 1.4$).  In this
way we are able to obtain maps 3.78$^\circ$ on a side, containing
$8192^2$ pixels, with a resolution of 1.66 arcsec. Notice
that the method to pile up boxes (sketched in Fig. \ref{fig:cubes}) can
introduce systematic errors on the calculation of the power spectrum
(see the corresponding discussion in Section \ref{sect:power_sp}).
We apply this method by varying the initial random seeds to produce 10
different light-cone realizations, used to assess the statistical
robustness of our results and the influence of cosmic variance.

To compute the signal produced by the SZ effects we need to convert
the line-of-sight integrals of equations (\ref{eq:ypar}) and 
(\ref{eq:bpar}) into expressions suitable for the \sph\ formalism. 
First, for the tSZ effect, we calculate the contribution of every gas 
element that lies inside the light-cone volume. We follow 
an approach similar to the one proposed by \cite{dasilva2000}. 
Given the $i$-th \sph\ particle, we define
\begin{equation}
\Upsilon_i \equiv \frac{k_B\, \sigma_T}{m_e\, c^2} \, N_{e,i} T_i \ ,
\label{eq:upsilon}
\end{equation}
where $T_i$ is the gas temperature and $N_{e,i}$ is the number of
electrons associated to the particle, i.e.
$N_{e,i}=n_{e,i}\,m_i/\rho_i$, being $n_{e,i}$, $m_i$ and $\rho_i$ the
electron density, the mass and the gas density of the $i$-th particle,
respectively.  Then, according to the required map resolution, we
compute the physical length of the pixel $L_{{\rm pix},i}$ at the
particle's distance from the observer and we use it to calculate
\begin{equation}
y_i \equiv \frac{\Upsilon_i}{L_{{\rm pix},i}^2} \ ,
\label{eq:y_i}
\end{equation}
which is the total contribution to the \ypar\ from the $i$-th
particle.  This quantity is then distributed over the map pixels by 
adopting the same \sph\ smoothing kernel which is used in the 
simulation code for the computation of hydrodynamical forces 
and proposed by \cite{monaghan1985}:
\begin{equation}
W(x)\propto \left\{ \begin{array}{ll}
1-6x^2+6x^3, & 0\le x<0.5\ , \nonumber \\
2(1-x)^3, & 0.5<x\le1\ , \\
0, & x>1. \nonumber
\end{array} \right.
\end{equation}
In the previous expressions $x\equiv\Delta\theta/\alpha_i$, where
$\Delta\theta$ represents the angular distance between the pixel
centre and the projected particle position and $\alpha_i$ is the angle
subtended by the particle smoothing length provided by the
hydrodynamical code.  In order to conserve the total intensity
associated to each particle we normalize to unity the sum of the
weights $W$ over all involved pixels.  Finally, the intensity in a
given pixel is obtained by summing over all the particles inside the
light-cone.

The construction of the maps of the Doppler \bpar\ follows an
identical procedure with the only difference that we have to
substitute equation (\ref{eq:upsilon}) and (\ref{eq:y_i}) with
\begin{equation}
B_i \equiv \frac{v_{r,i}}{c} \, N_{e,i} \ ,
\end{equation}
and 
\begin{equation}
b_i \equiv \frac{B_i}{L_{{\rm pix},i}^2} \ ,
\end{equation}
where $v_{r,i}$ is the radial component of the peculiar velocity of
the $i$-th particle, which can be either negative or positive.


\section{The distribution of the SZ signals}  \label{sect:signal}

\begin{figure*}
\includegraphics[width=0.42\textwidth]{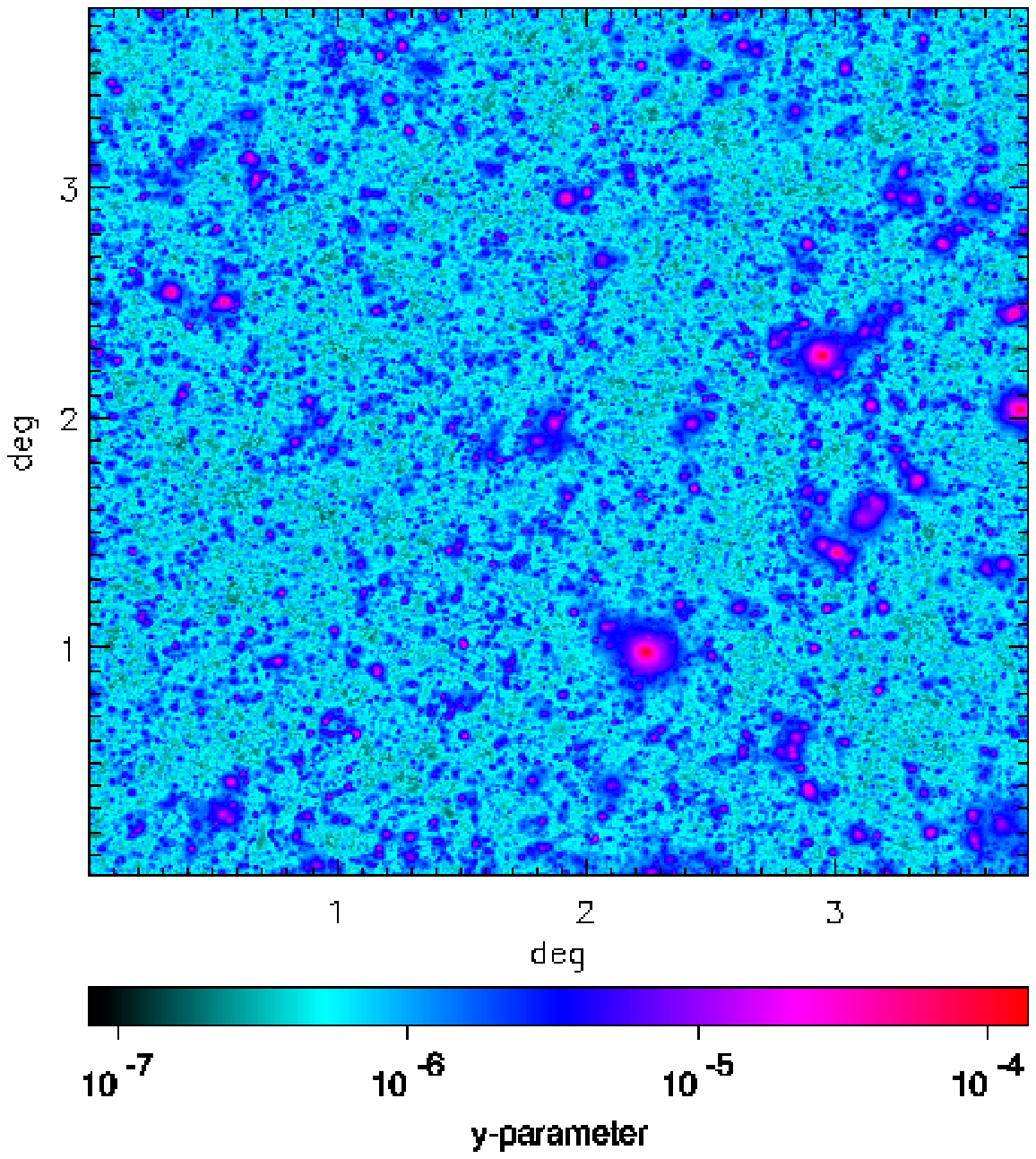}
\includegraphics[width=0.42\textwidth]{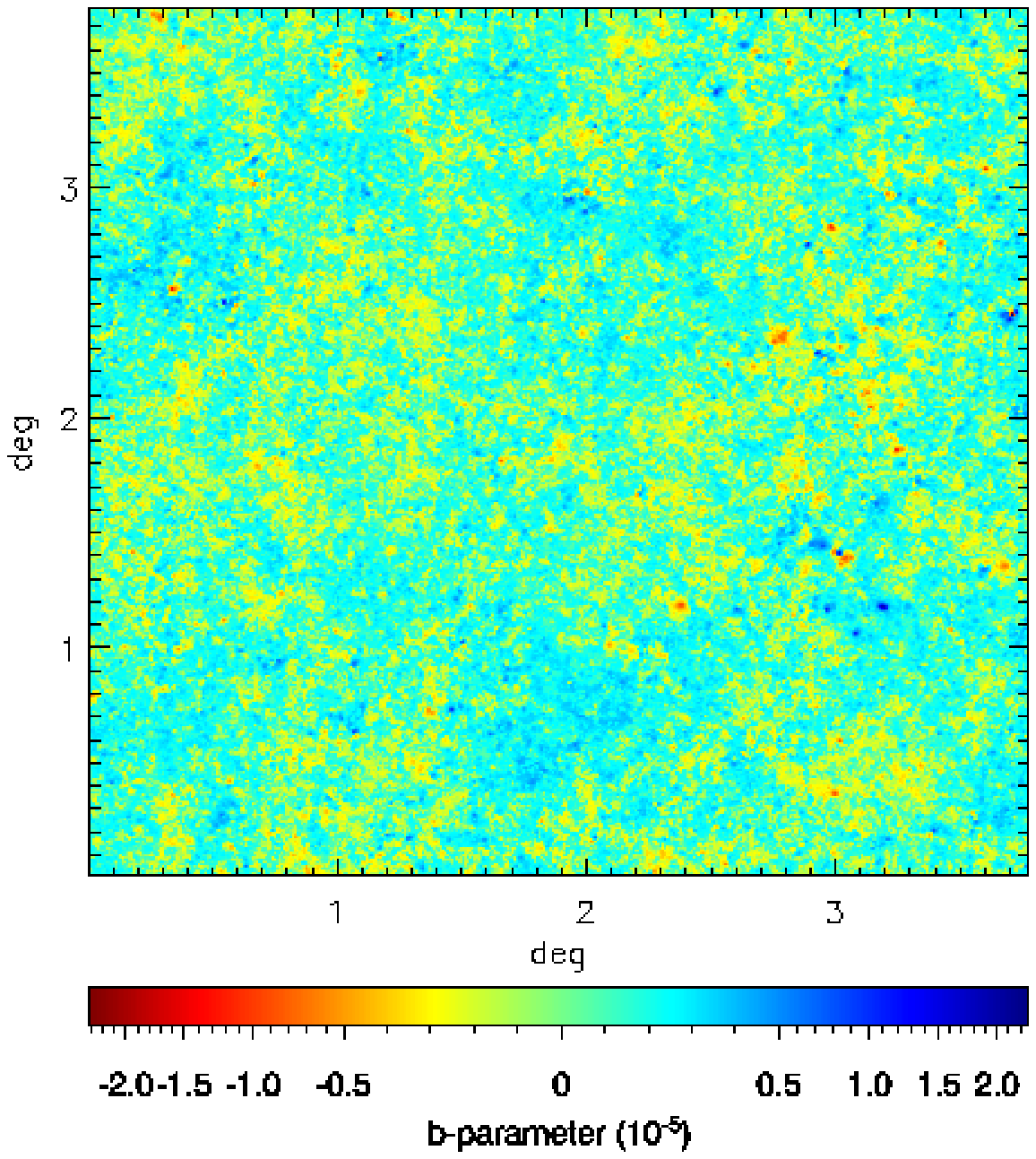} 
\caption{
Maps of the tSZ (left panel) and kSZ (right panel) signals expressed
in terms of the Compton \ypar\ and Doppler \bpar\ (given in units of
$10^{-5}$), respectively.  The maps are 3.78$^\circ$ on a side and the
pixel size is (1.66 arcsec)$^2$. Both maps refer to the same
realization of the past light-cone.  }
\label{fig:maps}
\end{figure*}

In Fig. \ref{fig:maps} we present two examples of the maps of the
Compton \ypar\ (left panel) and Doppler \bpar\ (right panel) obtained
with the method described in the previous section; both refer to the 
same light-cone simulation.

It is evident in the tSZ map that the signal is dominated by galaxy 
clusters that
can reach in their central regions typical intensities close to $y
\sim 10^{-4}$, which correspond to temperature changes an order of
magnitude higher than the primary CMB anisotropies, in the RJ limit.  
In the map it is
also possible to recognize a large number of smaller and fainter
structures with $y \sim 5 \times 10^{-6}$: they correspond to distant
protoclusters or more local galaxy groups. We also note a
non-negligible signal arising from regions outside clusters
(e.g. filaments and diffuse gas) that can reach at most $y \sim
10^{-6}$.  We use the whole set of maps to compute the expected
average of the Compton \ypar: we obtain a value of $<y>=(1.19 \pm
0.32) \times 10^{-6}$ as the mean of the pixel values in the 10 fields
of $3.78^2$ deg$^2$; the reported error represents the r.m.s. in
fields of 1 deg$^2$.  This value is significantly lower than the
results obtained by \citet[cooling model]{dasilva2001b} and
\citet[cooling plus feedback model]{white2002}: $2.3 \times 10^{-6}$
and $2.1 \times 10^{-6}$, respectively.  The main reason of this
discrepancy is the smaller value of $\sigma_8$ adopted in our
simulation ($\sigma_8=0.8$ against their $\sigma_8=0.9$): the mean 
value of the Compton \ypar\ is in fact expected to scale roughly 
as $\sigma_8^{\alpha/2}$, with $\alpha\approx 4-7$
\citep[see, e.g.,][]{sadeh2004,diego2004}.
For the same set of maps, we also repeat the analysis but
considering only the gas particles of the warm-hot intergalactic
medium \citep[WHIM, see][]{cen1999,dave2001}, defined as the gas
component having a temperature $T$ in the range $10^5$ $< T < 10^7$ K 
and we obtain $<y>_{\rm WHIM}=(6.90 \pm 0.42) \times 10^{-7}$, which 
means that about 60 per cent of the total value of the
\ypar \ is originated by WHIM gas, while the remaining fraction comes from
hotter gas ($T > 10^7$ K).

Analyzing the kSZ map, we notice that the peaks can reach values of $|b|
\gtrsim 10^{-5}$, i.e. roughly of the same intensity of the typical
$\Delta T/T$ of the primary CMB signal.  These peaks correspond to
clusters which are not necessarily massive and/or hot like for the tSZ
signal, but that have significant bulk motions. This leads to
considerable displacements in the peak positions in the two maps: for
example, we note that the brightest source in the tSZ map, roughly 
located in the position (2.2$^\circ$,1.0$^\circ$) and corresponding to 
a cluster 
at $z \sim 0.3$, is completely absent in the kSZ maps, since it is a 
relaxed object with a small radial component of the peculiar velocity.

\begin{figure*}
\includegraphics[width=0.42\textwidth]{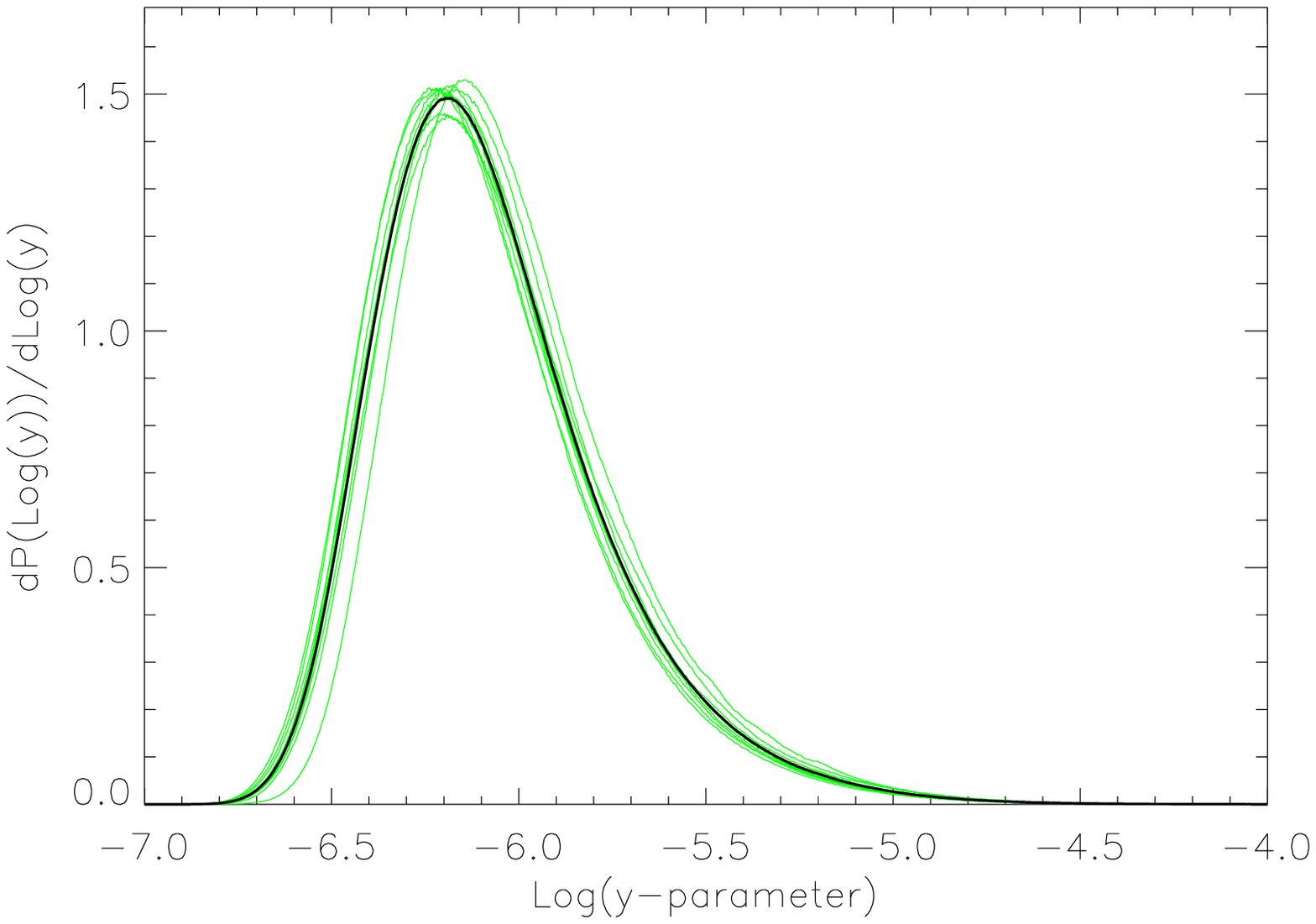}
\includegraphics[width=0.42\textwidth]{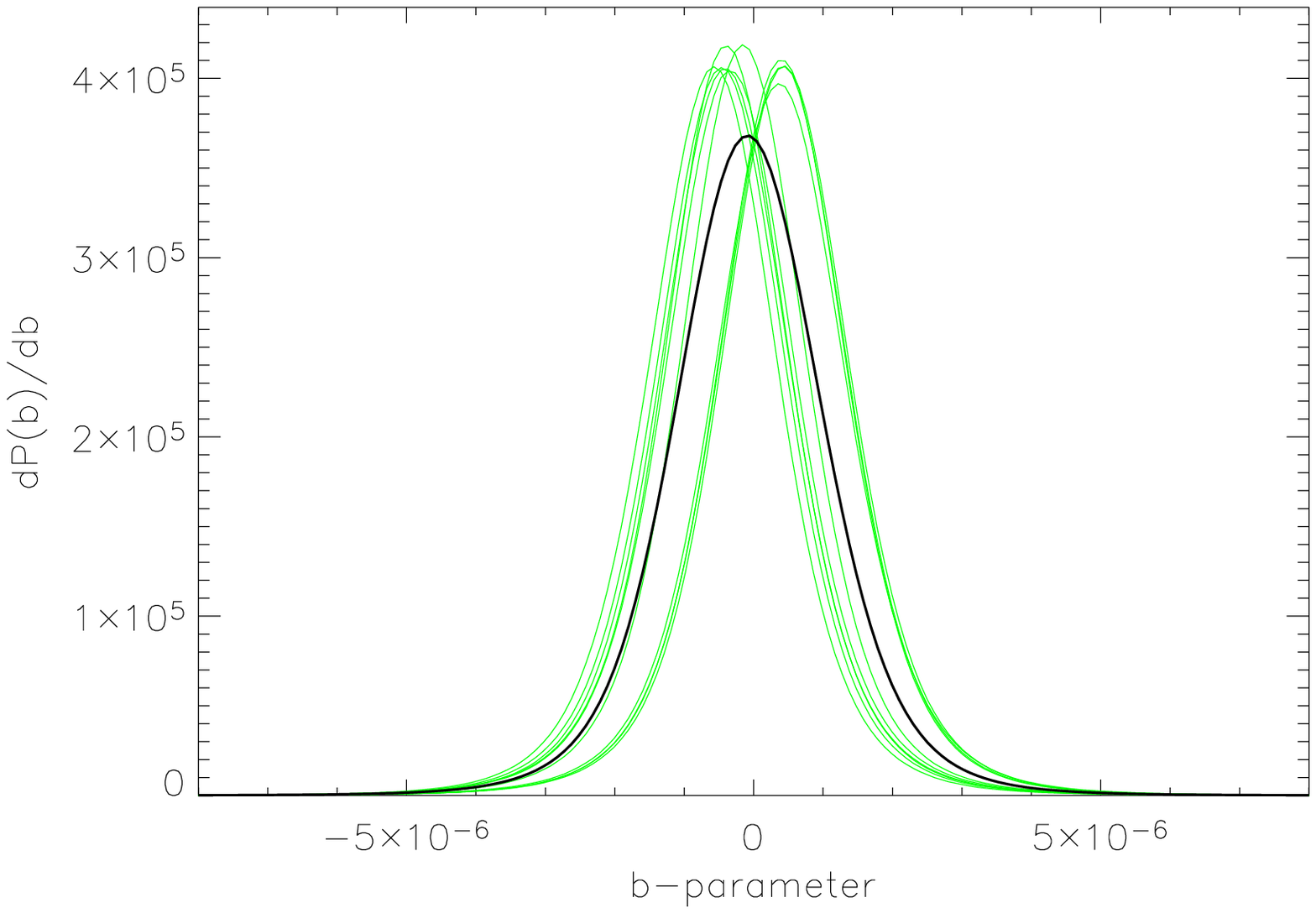}
\caption{
Probability distribution of the logarithm of the 
\ypar\ (left panel) and the \bpar\ (right panel), obtained considering the 
(1.66 arcsec)$^2$ pixels without beam smoothing.
The thin lines show the results for 10 different realizations; 
the thick solid line is the corresponding average.}
\label{fig:distr}
\end{figure*}

In Fig. \ref{fig:distr} we show the probability distribution function
of both SZ signals calculated considering the values in the (1.66
arcsec)$^2$ pixels: in particular the thin lines refer to each of the
10 light-cone realizations, while the thick line is the corresponding
average. We notice that the tSZ effect distribution is close to a 
lognormal function 
while the kSZ one is similar to a gaussian one.  Regarding the
average distribution of the tSZ signal we computed the first moments
of $I_y \equiv \log y$: we find $\overline{I_y}=-6.07$ with a
corresponding r.m.s. of 0.31 and a skewness\footnote{In this paper we
adopt the definition of skewness $s$ of a population $\{x_1,...,x_n\}$
given by the formula $s \equiv \frac{1}{(n-1)\sigma^3}\sum_{i=1}^n
(x_i-\overline{x})^3$, where $\overline{x}$ is the mean of the
distribution and $\sigma$ is its r.m.s.}  of 0.96. The odd moments of
the kSZ effect average distributions are close to zero, as expected, 
and we
obtain an r.m.s. of $1.3 \times 10^{-6}$ and a kurtosis\footnote{The
kurtosis $\kappa$ is defined as $\kappa \equiv
\frac{1}{(n-1)\sigma^4}\sum_{i=1}^n (x_i-\overline{x})^4-3 \,$.}  of
5.1.

\begin{figure*}
\includegraphics[width=0.42\textwidth]{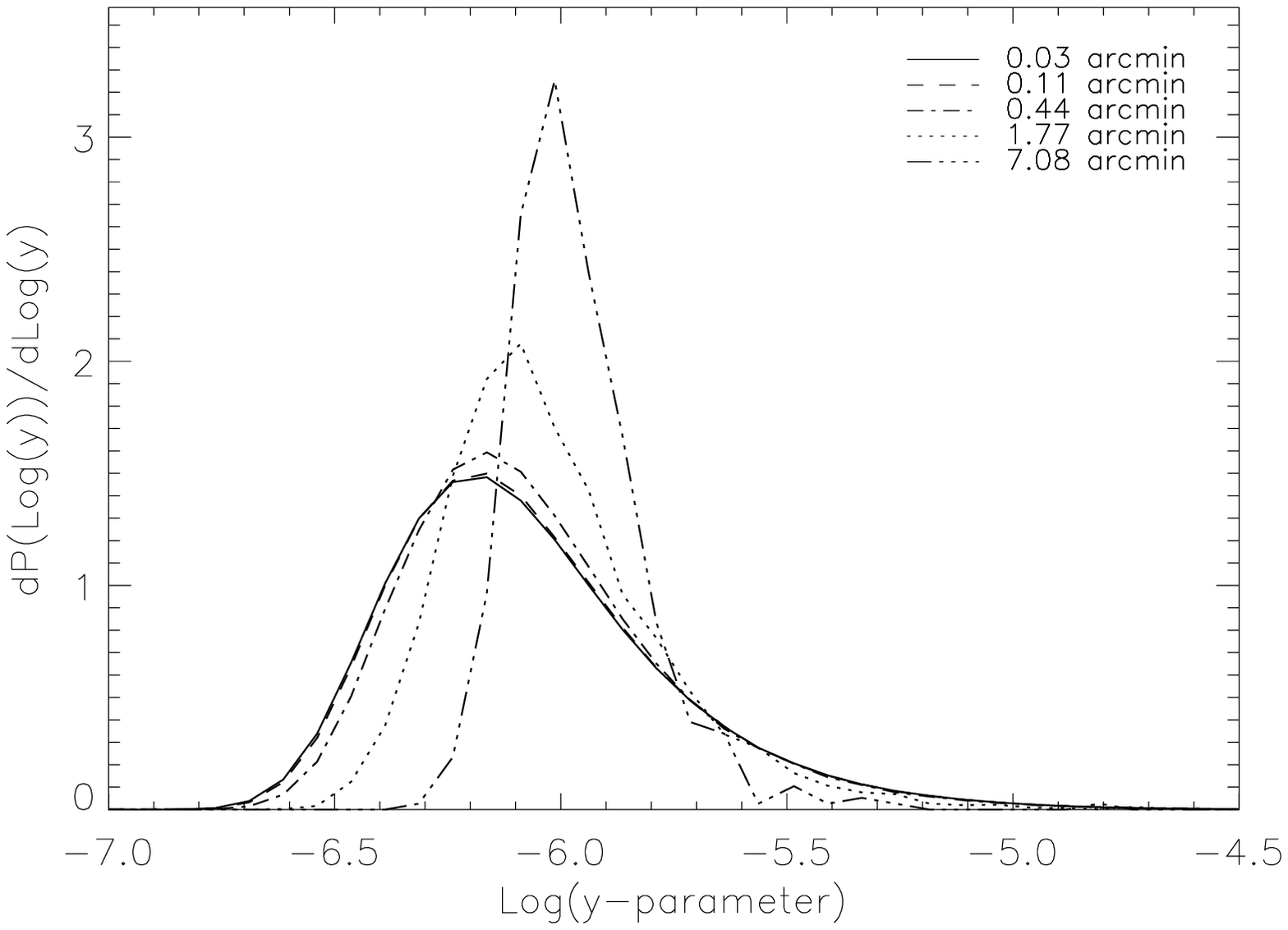}
\includegraphics[width=0.42\textwidth]{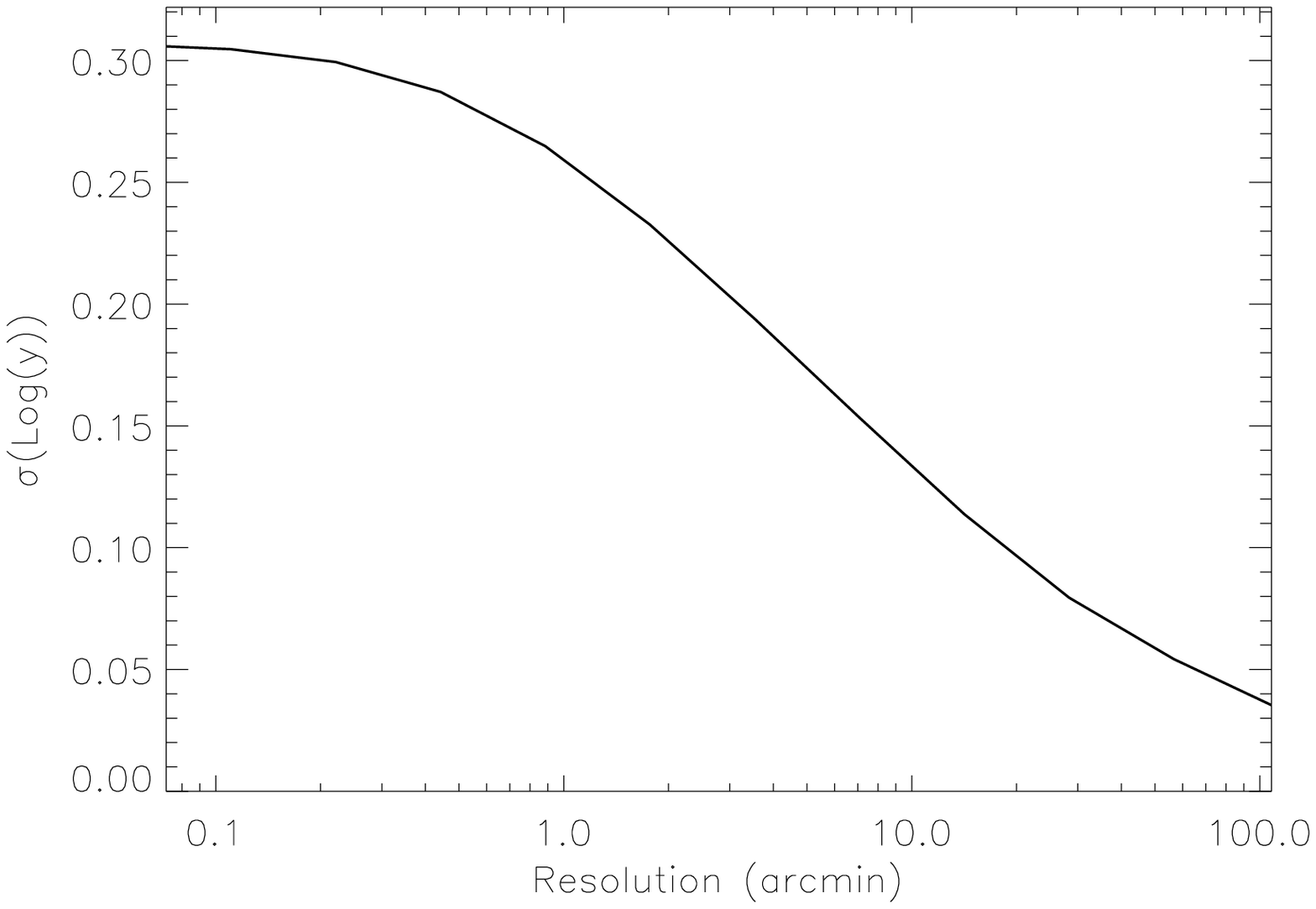}
\caption{Left panel: probability distribution functions of the 
logarithm of the \ypar\ in the 10 maps for different angular 
resolutions of the maps. Right panel: standard deviation of the 
distribution of the logarithm of the \ypar\ as a function of the 
beam smoothing.}
\label{fig:stat-Y}
\end{figure*}

We also analyse how these statistics change when we smooth our maps 
to simulate different instrument resolution. We show in the left panel of
Fig. \ref{fig:stat-Y} the dependence of the distribution of the
logarithm of the \ypar: the shape does not change significantly when smoothing 
the maps down to 0.44 arcmin (corresponding to $16^2$
pixels in the original map), while a further reduction of the
resolution leads to a change in the shape of the distribution with a
resulting absence of the lowest values in the map. On the contrary the
probability of obtaining high values remains unchanged also with a
beam smoothing of 1.77 arcmin, as a consequence of the fact that the peaks
correspond to galaxy clusters which have this typical angular size.
At low resolutions the distribution is also more peaked, as it can
also be seen in the right panel of Fig. \ref{fig:stat-Y}: the standard
deviation of the distribution drops from 0.3 to less than 0.1 when
varying the beam smoothing from 0.2 to 20 arcmin, indicating that in order
to capture the main features of the fluctuation field of the tSZ
effect a resolution better than $1$ arcmin is desirable.

\begin{figure*}
\includegraphics[width=0.42\textwidth]{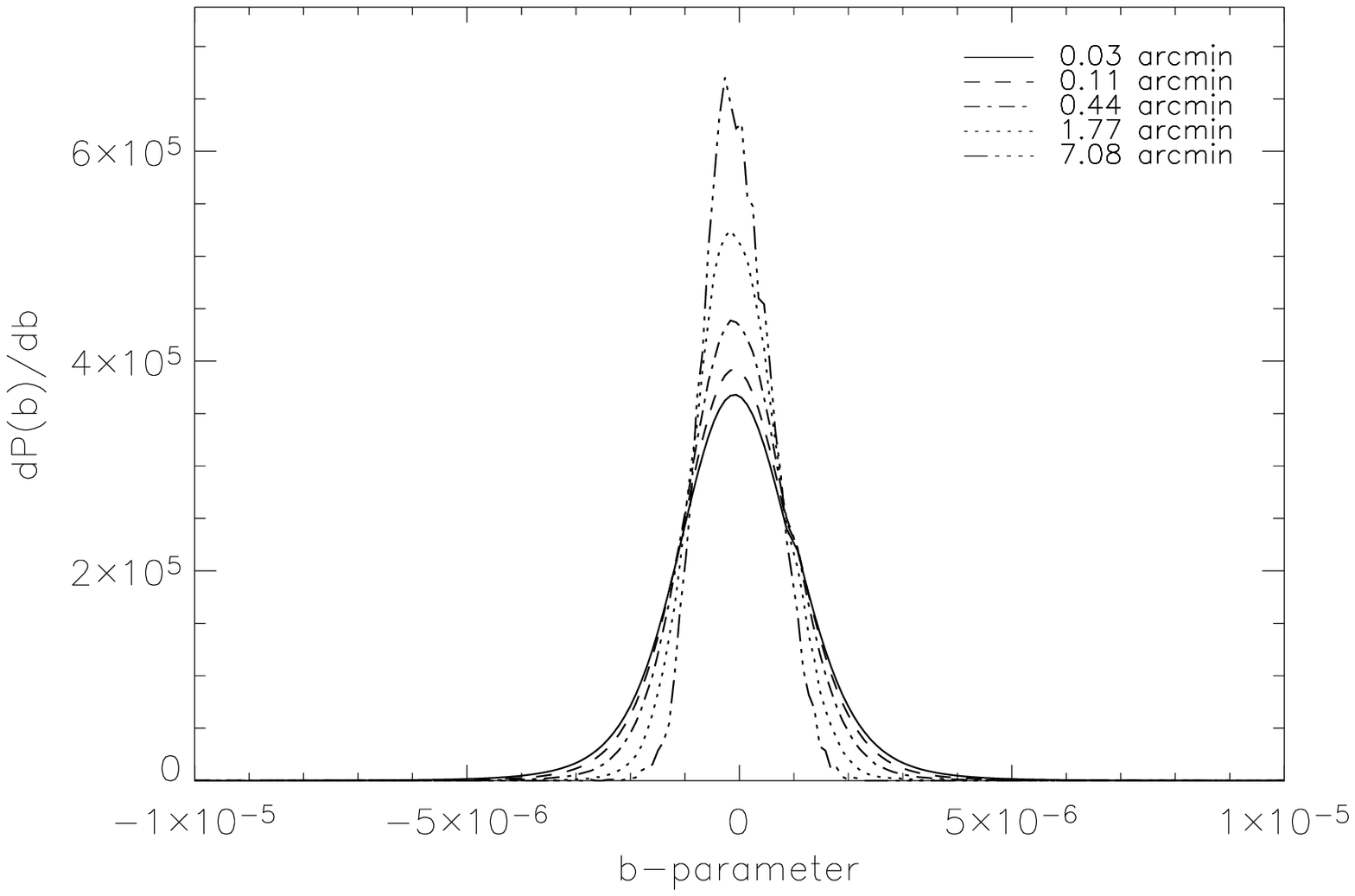}
\includegraphics[width=0.42\textwidth]{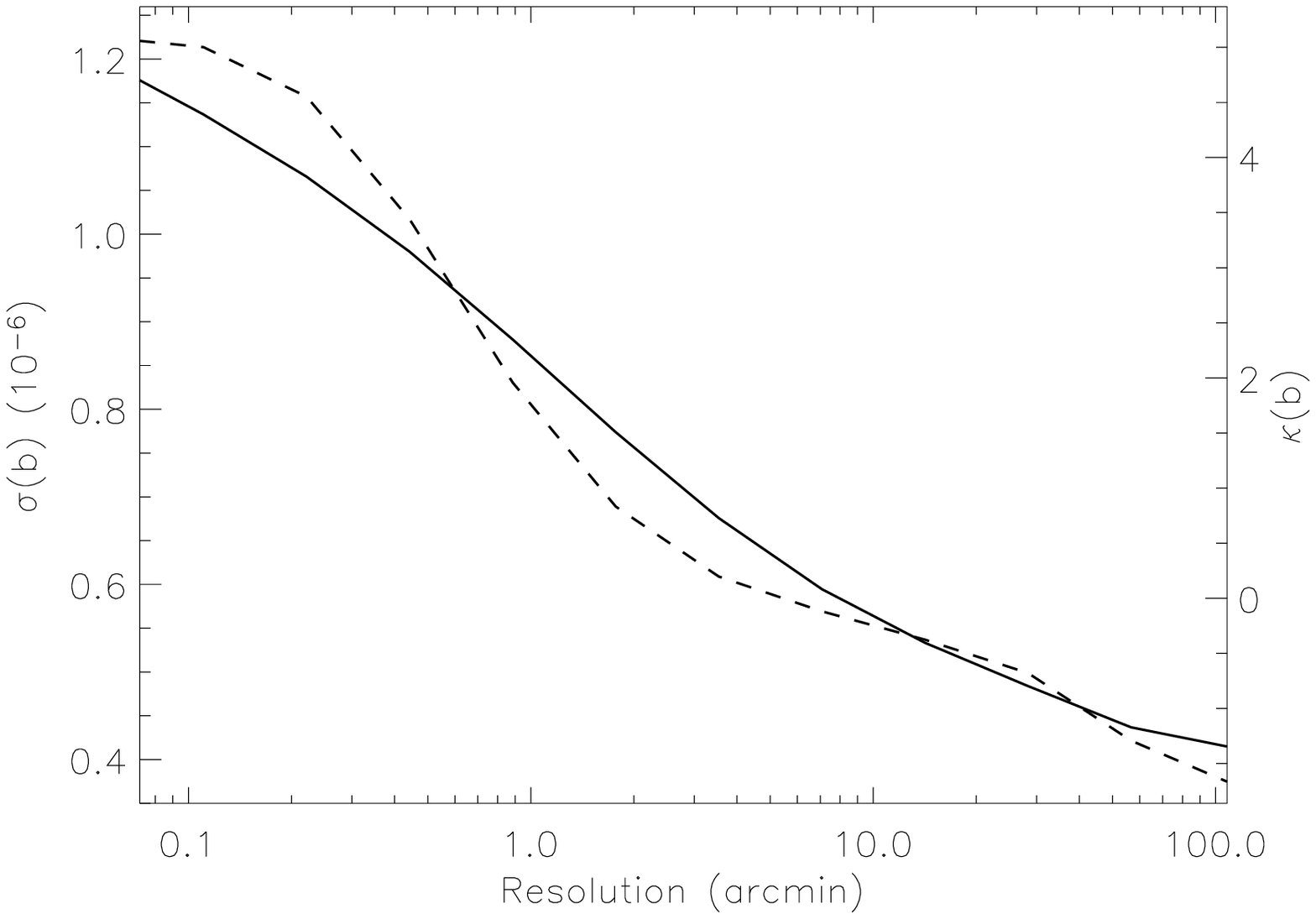}
\caption{Left panel: probability distribution functions of the  
\bpar\ in the 10 maps for different angular resolutions of the maps.
Right panel: standard deviation (solid line) and kurtosis (dashed line) of the 
distribution of the \bpar\ as a function of the beam smoothing.
}
\label{fig:stat-B}
\end{figure*}

We show in Fig. \ref{fig:stat-B} the corresponding plots for the 
distribution of the kSZ effect values. As expected the tails of the 
distribution 
function are strongly reduced by the beam smoothing 
\citep[our results are similar to those obtained by] []{dasilva2001a}. 
The right panel shows that the standard deviation (solid line)
increases at higher resolution due to the fact that, as we will
discuss more extensively in Section \ref{sect:power_sp}, the kSZ effect 
has a
significant contribution from small-scale signal. The level of
non-gaussianity of the distribution can be expressed in terms of the
kurtosis $\kappa$ also plotted in the right panel of
Fig. \ref{fig:stat-B} (dashed line). The distribution of values is
very close to a gaussian (i.e. $\kappa =0$) for resolutions of $\sim$1
arcmin, while at higher resolution the kurtosis increases up to more
than 4 thus populating more the tails of distribution. This indicates
that the non-gaussianity originated by the kSZ effect must be taken
into account when measuring the primordial (i.e. inflationary)
non-gaussianity in the CMB signal at scales lower than 1 arcmin.

\begin{figure*}
\includegraphics[width=0.42\textwidth]{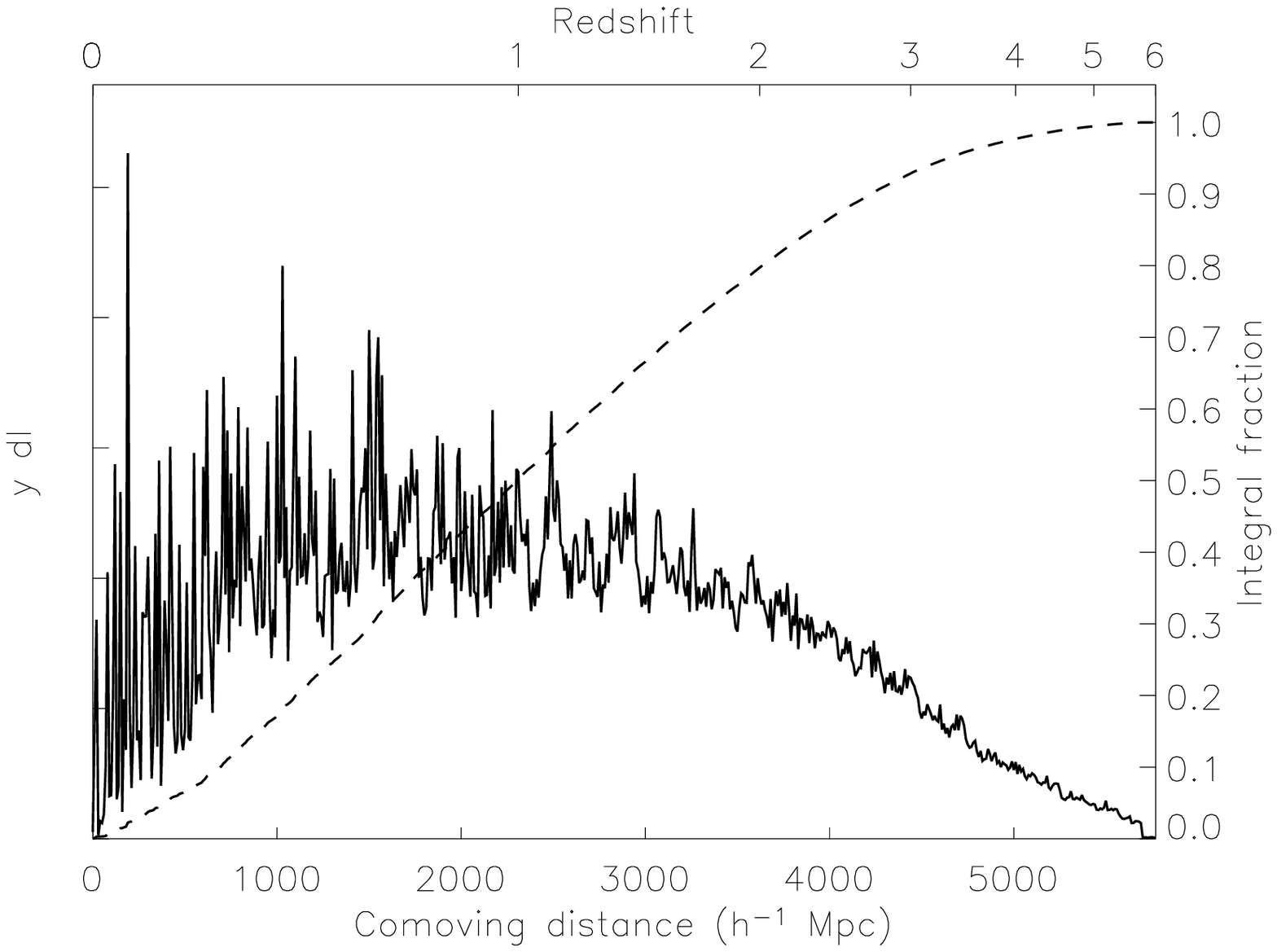}
\includegraphics[width=0.42\textwidth]{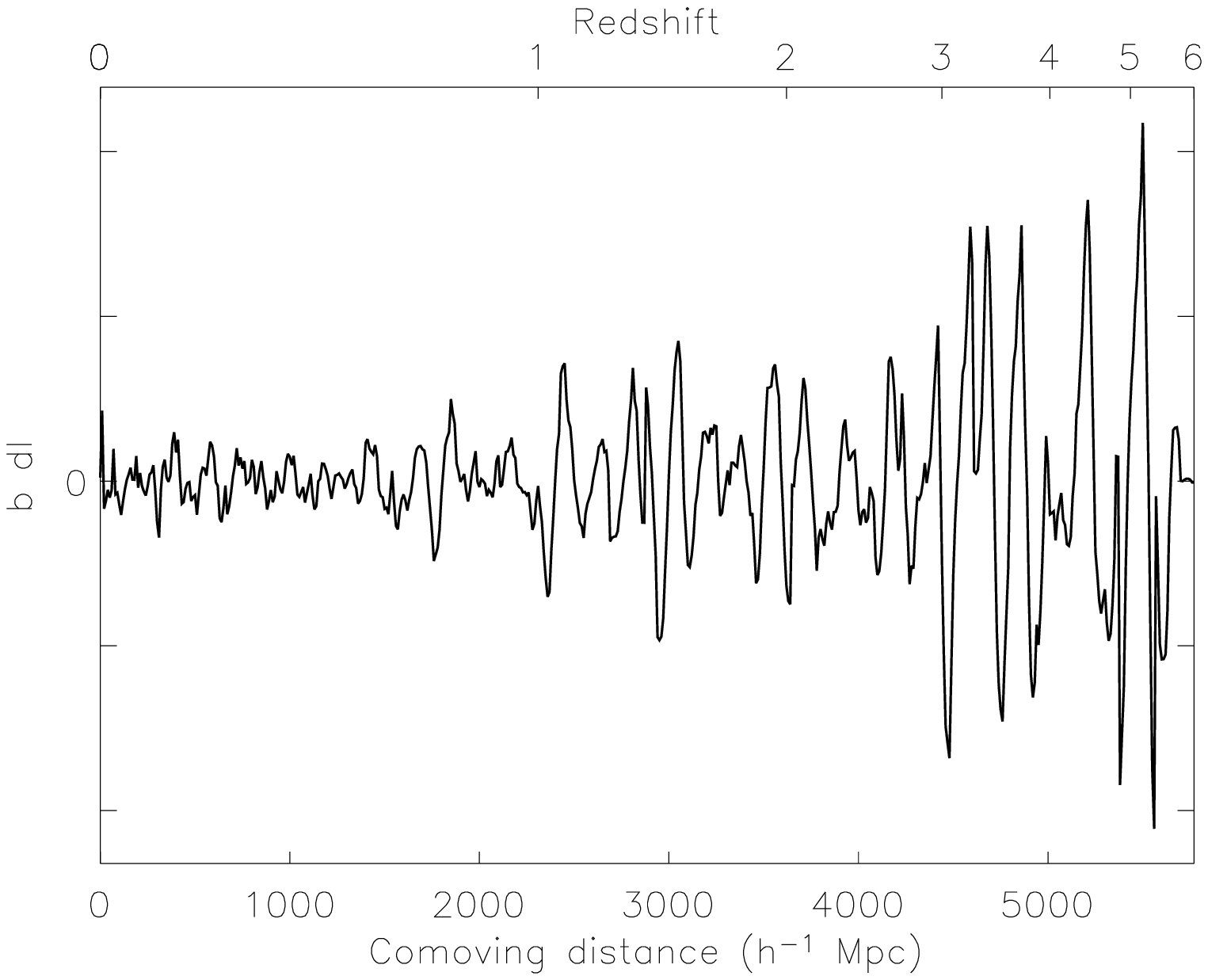}
\caption{
Differential contributions to the \ypar\ (left panel) and \bpar\
(right panel) as a function of the comoving distance from the
observer; the corresponding redshifts are indicated on the top.  The
values, representing the mean of the 10 map realizations, are computed
in equispaced comoving distance intervals with length of $10 \,
h^{-1}$ Mpc. The dashed line in the left panel shows the integral of 
the distribution as a fraction of the final average value of the \ypar.}
\label{fig:zdistr}
\end{figure*}

As already mentioned in Section \ref{sect:simulation}, unlike the
X-ray signal, both the tSZ and the kSZ effects have significant
contributions from high-redshift gas.  This can be clearly seen in
Fig. \ref{fig:zdistr}, where we show the contribution to the \ypar\
and \bpar divided into equal comoving distance intervals with length
of $10 h^{-1}$ Mpc (the corresponding redshifts are indicated on the
top of the panel). Note that the curves represent the average value of
the 10 different light-cone realizations. The tSZ signal (left panel)
shows several spikes at low redshift ($z<0.5$), mainly due to the
presence of galaxy clusters in the bin. At higher redshifts the
distribution is much more regular because the dispersion becomes
lower, since the light cones include larger comoving volumes at larger 
distances. Moreover large collapsed structures are very rare events at
$z \gtrsim 2$.  Integrating this redshift distribution we obtain, in
good agreement with previous analyses \citep[see,
e.g.,][]{dasilva2000}, that half of the total tSZ signal comes from
$z>1$, and only about 20 per cent from $z>2$.

Since in a given redshift bin there are both gas elements approaching
and receding from the observer with similar probabilities, the kSZ
signal (right panel), which is the algebraic sum of all the
contributions, has a vanishing expectation value.  Therefore the
distribution plotted in the right panel of Fig. \ref{fig:zdistr} can
be considered as a measurement of its dispersion. The plot clearly
shows that, even at very high redshift, the contribution to the kSZ 
effect is
non-negligible: as suggested by \cite{dasilva2001a}, the increase of
the contributing gas mass compensates the decrease of its
velocities.


\section{The SZ angular power spectra} \label{sect:power_sp}

\begin{figure}
\includegraphics[width=0.45\textwidth]{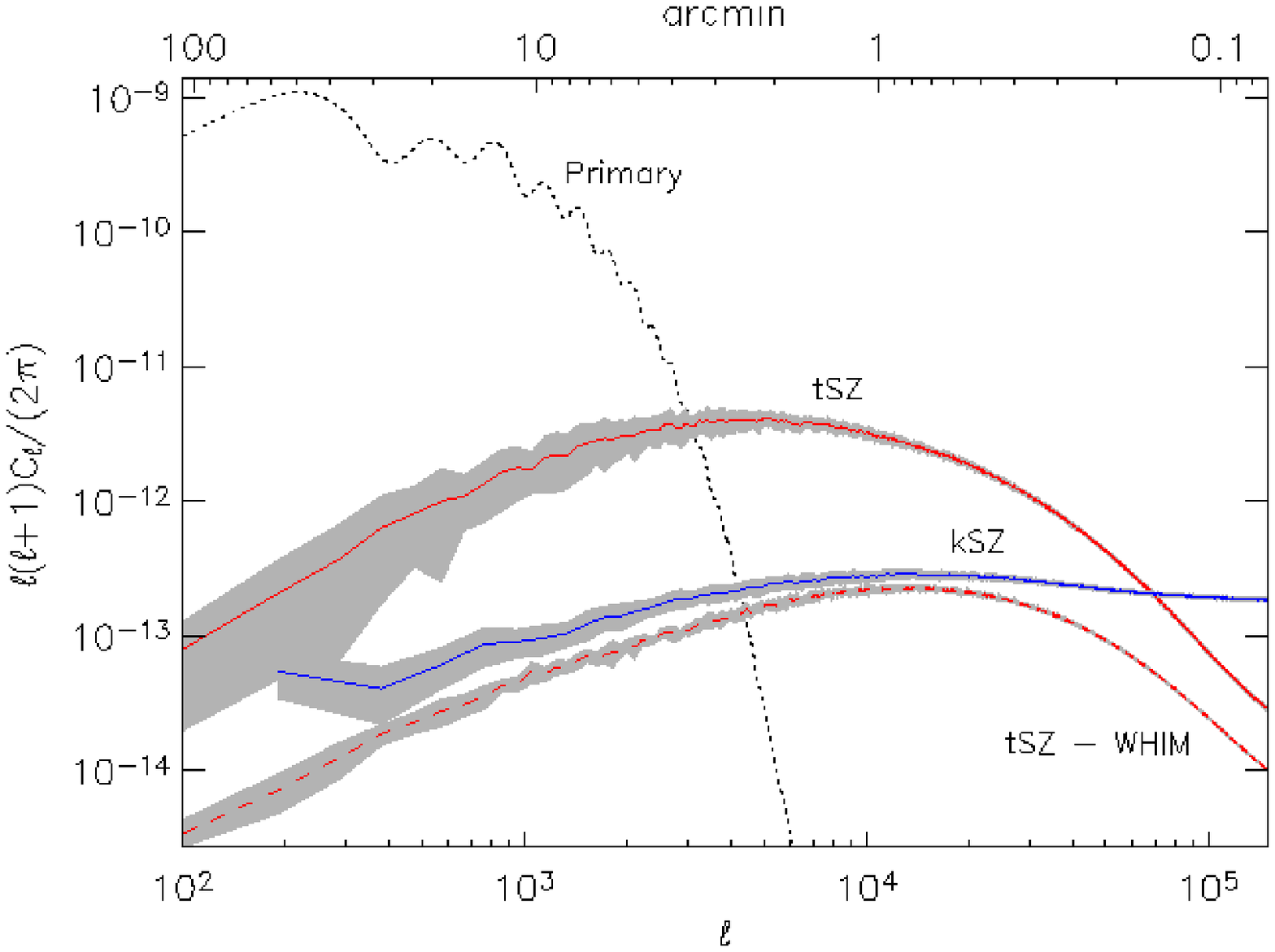}
\caption{
Angular power spectra of the $\Delta T/T_{\rm CMB}$ induced by the
different SZ effects as a function of the multipole $\ell$. The dotted
line represents the primary CMB signal calculated using
\textsc{cmbfast} \protect \citep{seljak1996} and assuming the same
cosmological model considered for our hydrodynamical simulation.  The
solid lines are the tSZ and kSZ power spectra: they represent the
average of the power spectra of different maps (10 light-cones for the
tSZ effect, 40 for the kSZ one).  The dashed line refers to the tSZ power
spectrum for the WHIM. The shaded regions represent the
r.m.s. calculated between the 10 (40) maps.  The tSZ spectra are
computed at the frequency $\nu =$ 30 GHz.}
\label{fig:pow_sp}
\end{figure}

In order to study the detectability of the SZ effects, it is important
to analyze not only the intensity of the signal itself, but also its
power at different angular scales.  In Fig. \ref{fig:pow_sp} we
compare the angular power spectra for both SZ effects to the primary
CMB one.  The displayed SZ spectra, which represent the average of
the power spectra of the 10 different light-cone realizations, have been
obtained in the approximation of flat sky (which well holds for $\ell >
100$) and using a method based on Fast Fourier Transform.
We considered a frequency $\nu = 30$ GHz for the tSZ effect, that 
corresponds to
$g_\nu(x)=-1.94$ (see equation \ref{eq:g_nu}), thus near the RJ limit for
which the effect is maximum, while the choice of the frequency does not 
affect the intensity of the kSZ signal (see Section \ref{sect:sze}). 
 As already discussed in Section
\ref{sect:models}, both the SZ effects have a significant contribution 
from high-redshift gas. Since our map-making algorithm 
replicates the same volume four times at $z>1.4$ (see
the comments on Fig. \ref{fig:cubes}), this could introduce a 
spurious correlation in the power spectrum calculations, 
therefore, we need to perform some checks to assess the reliability of 
our results. Our analysis indicates that the kSZ maps are affected 
by this
problem: in fact the kSZ power spectrum computed with this method 
presents some ``fake'' peaks at high angular scales generated by 
gas located at $z>1.4$: for this reason we decided to split each 
of the 10 kSZ maps into 4 submaps (thus creating 40 different light 
cones) and to calculate their power spectra separately. We also 
checked that the tSZ maps are not affected by this problem (e.g. the 
power spectra calculated with maps and submaps are identical), 
therefore for the tSZ effect we keep the full maps.

Given the size and resolution of our maps, in principle this 
procedure can allow us
to compute the power spectra from $\ell \simeq 95$ ($\ell
\simeq 190$ for the kSZ effect) out to $\ell \simeq 7.8 \times 10^5$. 
However the finite box size of 192 $h^{-1}$ Mpc of our simulation makes 
our results reliable only for $\ell \gtrsim
200$ (e.g. the angular extension of 1/2 of the box size at $z=6$). 
Anyway for the purpose of this paper, we are not interested in
investigating the properties of the SZ signal at $\ell \lesssim 1000$, 
because in this range the primary CMB anisotropies dominate and because 
at large scales the tSZ signal is highly affected by the local structures 
\citep[see, e.g.][]{dolag2005b}. For what concerns the 
small angular scales, our full resolution maps resolve the 
gravitational softening of the simulation for distances from the 
observer lower than $\sim1000 \, h^{-1}$ Mpc: therefore, considering 
also the spatial resolution imposed by the \sph\ code, we can consider 
our results reliable only for resolutions higher than 3-4 arcsec 
($\ell \simeq 2 \times 10^5$), that subtend distances higher than the 
gravitational softening for $z \gtrsim 0.17$, from which most of the 
SZ signals arise, as discussed in Section \ref{sect:signal}.

The tSZ power spectrum peaks at $\ell \simeq 5000$ and, in the RJ
regime, it starts to dominate the primary CMB signal at scales of
about 4 arcmin. The kSZ signal is about one order of magnitude lower
at these scales. However, it is interesting to note that while at higher 
$\ell$ the tSZ signal looses power
significantly, our model predicts an almost flat kSZ power spectrum,
that overcomes the tSZ one for $\ell \gtrsim 7 \times 10^4$ even in the 
RJ regime. This behaviour, which is similar to the one obtained by 
\cite{zhang2004}, is a consequence of both the high contribution from 
distant, and therefore smaller, objects that affect the kSZ effect
signal and of the effect of the galactic winds present in our
simulation that are able to increase the power of the signal on very
small scales.  The tSZ signal arising from the WHIM (also plotted in
Fig. \ref{fig:pow_sp}) peaks at the scales of about 1 arcmin, which
roughly corresponds to the angular scales of galaxy groups. Even if
the total signal of the WHIM contributes to about 60 per cent of the
mean value of the \ypar\ (as we already noted in Section
\ref{sect:signal}), at the angular scales where the tSZ signal is
dominant the amplitude of its power spectrum contributes to less than
10 per cent of the total tSZ one: this shows that the detection of the
tSZ signal constitutes a probe almost only for the ICM at temperatures
$T > 10^7$ K, that mainly resides in collapsed structures.

\begin{figure}
\includegraphics[width=0.45\textwidth]{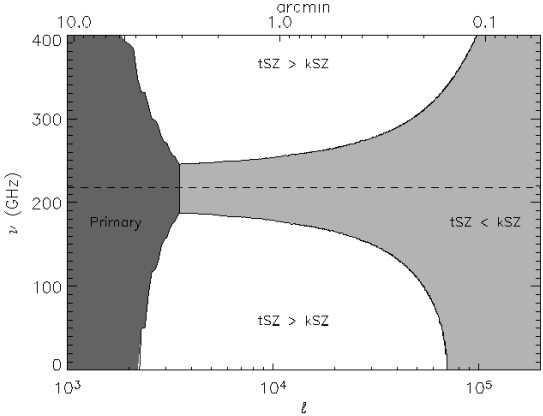}
\caption{
In the $\nu-\ell$ (observed frequency-angular scale) plane we show the
regions where the power spectra corresponding to different signals are
dominant.  Dark shaded, light shaded and white regions show where the
primary CMB signal, the kSZ effect and the tSZ effect dominate the
other components, respectively. The solid lines show where two effects
have the same power, the dashed line corresponds to the frequency of
218 GHz, at which the tSZ signal vanishes.}
\label{fig:freq_l}
\end{figure}

As already noted in Section \ref{sect:sze}, the intensity of the kSZ
effect does not depend on the observed frequency. On the contrary,
when increasing the frequency in the RJ regime, the tSZ effect
decreases, so the angular scale at which the kSZ signal dominates over
the tSZ one becomes higher. Fig. \ref{fig:freq_l} shows the dependence 
on the angular multipole of the frequency of equality between the 
power spectra of the two effects.  It is worth to note that at 150 GHz 
the kSZ effect dominates for $\ell \gtrsim 3 \times 10^4$.

\section{Cross-correlation between the SZ signals}\label{sect:cross}

\begin{figure}
\includegraphics[width=0.45\textwidth]{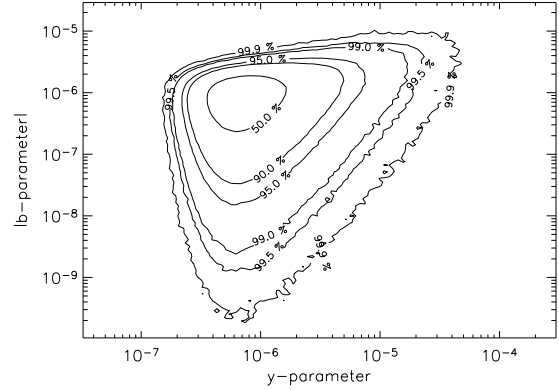}
\caption{Distribution of the pixel values 
for the whole set of 10 maps: the (modulus of the) Doppler \bpar\
vs. the Compton \ypar.  The 6 contour levels enclose 50, 90, 95, 99,
99.5 and 99.9 per cent of the total amount of pixels, respectively.}
\label{fig:sc_y-b}
\end{figure}

The two SZ signals both depend linearly on the electron density $n_e$
and receive the main contribution from the same cosmic structures, so
a correlation between the two is expected. However, since temperature and
velocity of the gas are not tightly correlated and since the \bpar\
depends only on the radial component of the velocity, a spread is
expected also. This can be seen in the contour plot in
Fig. \ref{fig:sc_y-b} that represents the distribution of pixel values
according to the two SZ effects: a significant scatter is present for
low ($y \lesssim 10^{-6}$) values of the \ypar\ to which can
correspond \bpar\ from zero almost to the highest possible
values. Anyway, considering the contour levels, most of the surface of
the sky is expected to have average values of the two signals.

\begin{figure}
\includegraphics[width=0.45\textwidth]{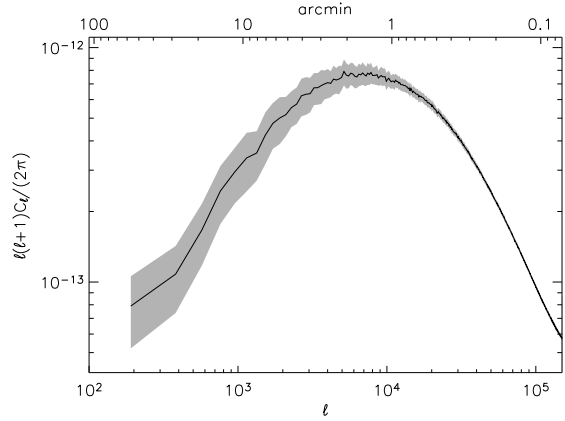}
\caption{
Power spectrum of the cross-correlation tSZ-kSZ 
at $\nu=30$ GHz as a function of the multipole $\ell$. The solid 
line represents the average of the power spectra of the 40 submaps;
the shaded region shows the corresponding r.m.s.
}
\label{fig:cps_y-b}
\end{figure}

We show in Fig. \ref{fig:cps_y-b} the power spectrum of the
cross-correlation between the two SZ signals computed at 30 GHz 
(the scaling with the frequency $\nu$ is simply given by $g_\nu(x)$ 
of equation \ref{eq:g_nu}). The correlation peaks at 2 arcmin, the typical 
scale of galaxy clusters, and a has sharp decrease at higher $\ell$ 
due to the lack of power of the tSZ effect.

\begin{figure}
\includegraphics[width=0.45\textwidth]{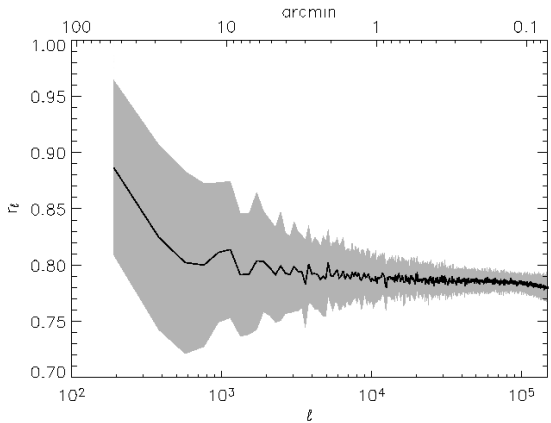}
\caption{
Cross-correlation coefficient between the tSZ and the kSZ effects as a
function of the multipole $\ell$. The solid line represents the
average of the correlation coefficients of the 40 submaps; the shaded
region shows the r.m.s. between them.  }
\label{fig:rl_y-b}
\end{figure}

In order to quantify the strength of the correlation we follow 
\cite{cheng2004} and compute the cross-correlation coefficient 
between the two signals which is defined as
\begin{equation}
r_\ell \equiv \frac{C_\ell^{\rm tSZ-kSZ}}{ \sqrt{C_\ell^{\rm tSZ}C_\ell^{\rm kSZ}} } \ .
\label{eq:r_l}
\end{equation}
The corresponding results are shown in Fig. \ref{fig:rl_y-b}. Globally
the tSZ and kSZ effects present a high correlation ($r_\ell \simeq 0.78$),
almost independent of the angular scale, indicating that, as already
said in the comments on Fig. \ref{fig:sc_y-b}, the average properties
of the two signals are similar despite of the different physical
dependence. The spread of the correlation in the different light-cones
(indicated by the shaded region) is also quite low ($\sim 0.01$); at
low $\ell$ it increases ($\sim 0.07$) due to the lack of statistics
and the larger cosmic variance.

\section{Cross-correlation with the soft X-ray signal}\label{sect:X-ray}

The ionized gas responsible of the SZ signal also produces an X-ray
emission. However, the dependence on both density/scale
and redshift is different. Consequently, the cross-correlation between the two 
signals can provide information on the scales, masses and redshifts 
contributing to each signal.

\begin{figure*}
\includegraphics[width=0.45\textwidth]{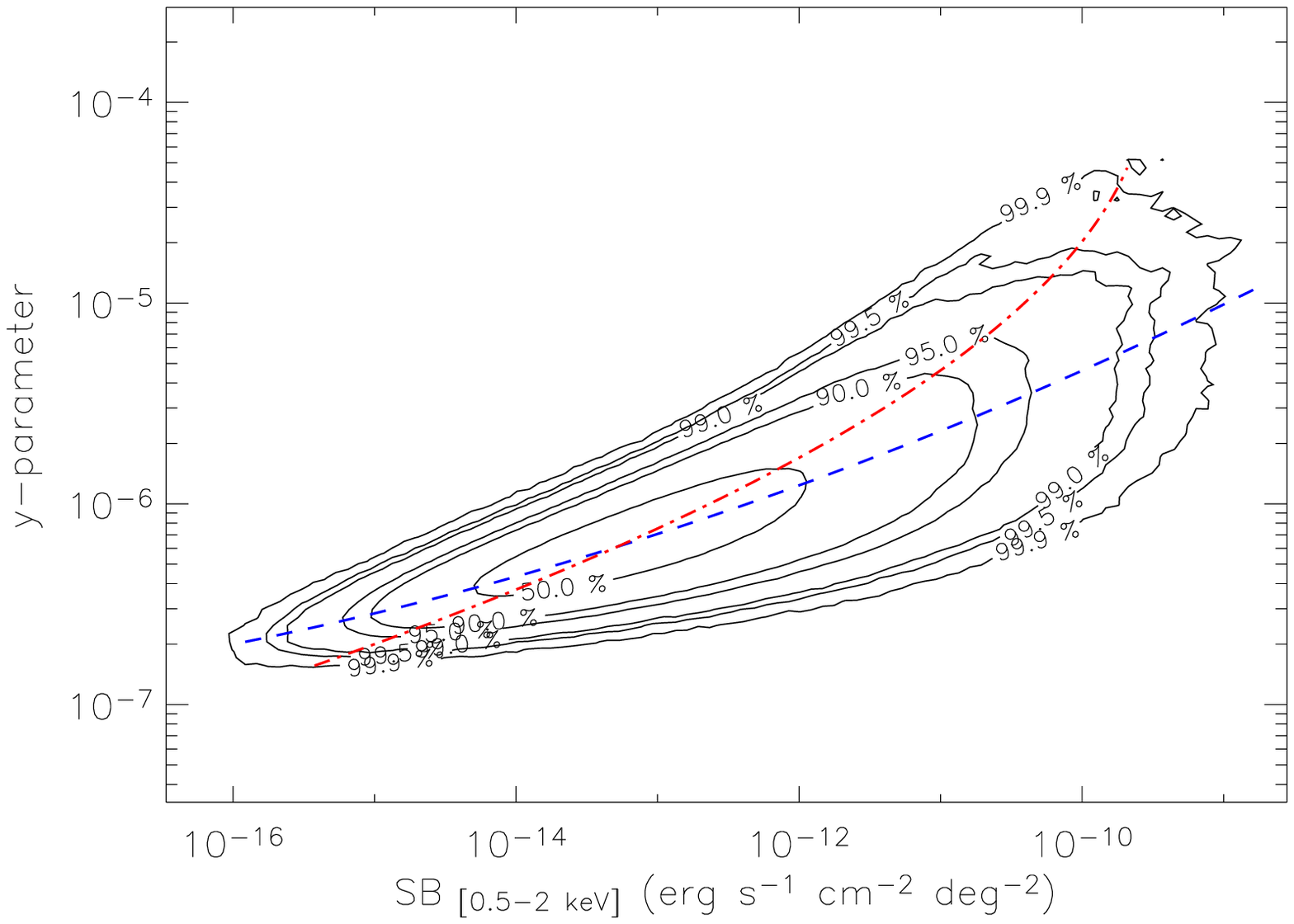}
\includegraphics[width=0.45\textwidth]{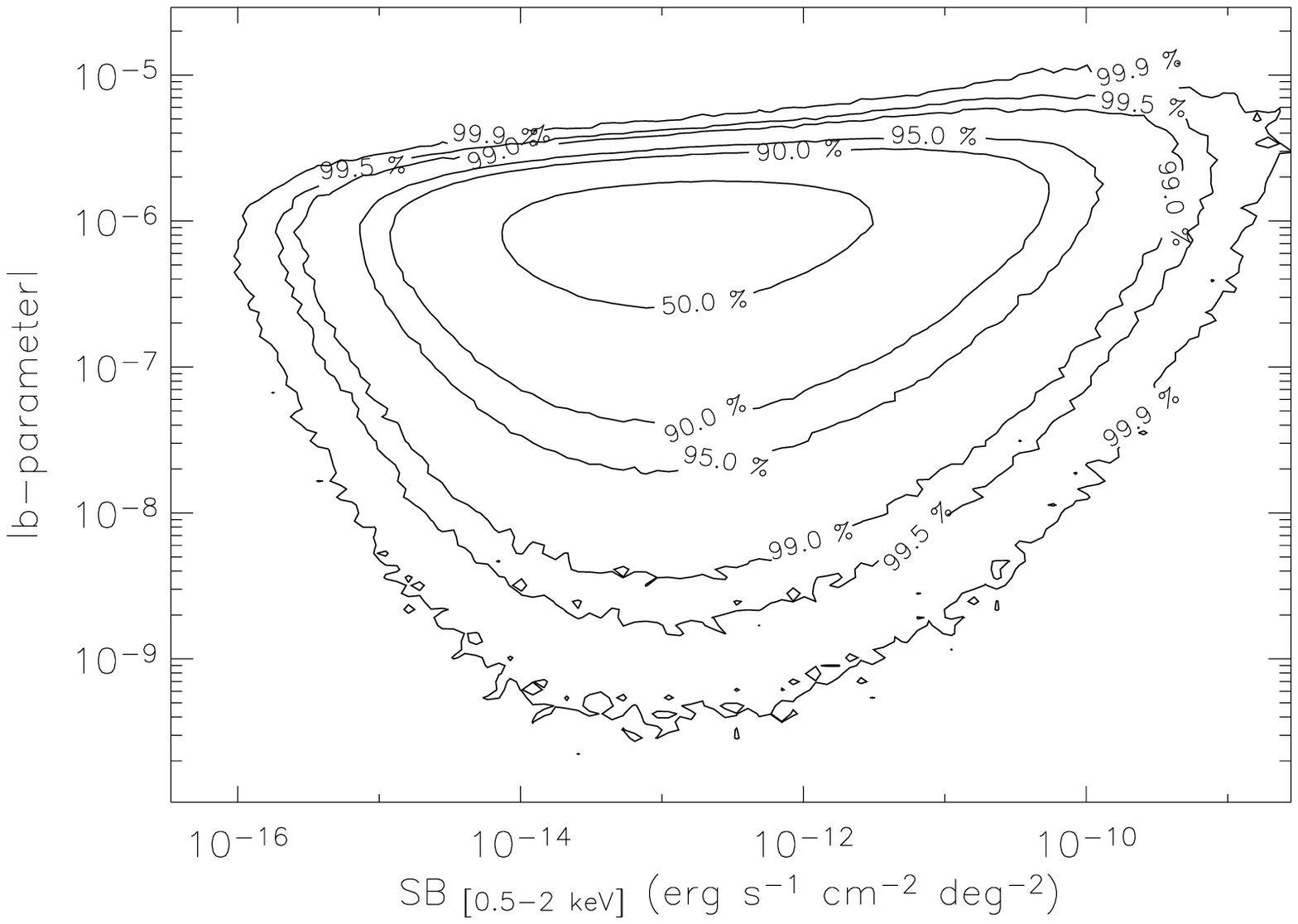}
\caption{Left panel: distribution of the pixel values 
for the whole set of 10 maps:
the Compton \ypar\ vs. the soft (0.5--2 keV) X-ray surface 
brightness. The 6 contour levels enclose 
50, 90, 95, 99, 99.5 and 99.9 per cent of the total 
amount of pixels, respectively. The dashed line represents the maximum-likelihood 
relation $Y(X)$, obtained by fitting the relation in equation 
(\ref{eq:fit_par}); the dot-dashed line represents the opposite 
maximum-likelihood relation $X(Y)$. Right panel: as the left panel,
but for  the (modulus of the) Doppler \bpar\ vs. the soft (0.5--2 keV) 
X-ray surface brightness.
}
\label{fig:sc_soft-yb}
\end{figure*}

The contours in the left panel of Fig. \ref{fig:sc_soft-yb} represent 
the distribution of pixel values according to the soft (0.5--2 keV) X-ray 
surface brightness (SXRB) and the tSZ effect: it is evident that most 
of the regions in the sky present both a low soft X-ray emission 
($\sim 10^{-14}-10^{-12}$ \sbunits) and 
\ypar\ ($\sim 10^{-6}$), as shown by the distributions reported in 
Fig. \ref{fig:distr}. We also notice that, as expected, the amplitudes 
of the two signals are correlated, even if a significant scatter is 
present especially for high intensities. Using this 
bi-dimensional distribution we calculate the maximum-likelihood 
relation that for a given value of the X-ray surface brightness 
provides the corresponding one of the \ypar. We proceed in the 
following way: given $X \equiv 
\log_{10}($SB(\sbunits) $)$ and $Y \equiv \log_{10}(y)$, we consider 
separately the distribution associated to all the different values of 
$X$ (i.e. the columns of the matrix shown in the left panel of Fig. 
\ref{fig:sc_soft-yb}); then for a given $X$ we take the value of $Y$ 
corresponding to the maximum point of the distribution and its 
scatter and we fit these data with a polynomial relation given by 
the formula
\begin{equation}
Y = a_0 + a_1 X + a_2 X^2 \ ,
\label{eq:fit_par}
\end{equation}
where and $a_{i=0,2}$ are the free parameters. We also weight the data  
according to the total amount of points of each distribution. We obtain 
the best-fit relation of $Y = -0.72 + 0.61X + 0.015X^2$ represented in 
the plot by the dashed line. Switching the two variables of equation 
(\ref{eq:fit_par}) and with the same procedure we obtain also the 
opposite relation $X = -21 -5.8Y -0.73Y^2$ shown by the dot-dashed 
line.

The right panel of Fig. \ref{fig:sc_soft-yb} shows the scatter plot
between the soft X-ray surface brightness and the (modulus of the)
\bpar. The dependence of the kSZ effect from the X-ray emission is
weaker than the tSZ effect one, however this plot shows that the
highest peaks of the kSZ effect ($|b| > 10^{-5}$) are always
associated to high surface brightness values (e.g. inner regions of
galaxy clusters).

\begin{figure}
\includegraphics[width=0.45\textwidth]{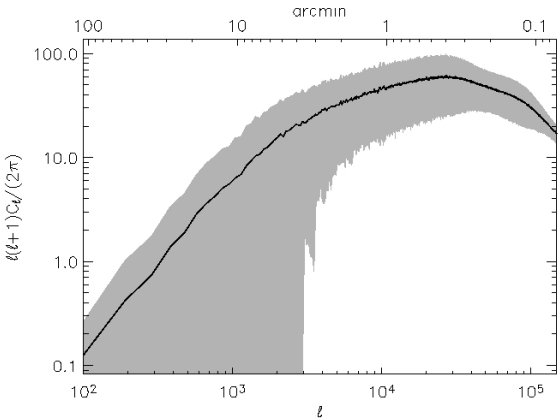}
\caption{
Power spectrum of the SXRB in the soft (0.5--2 keV) X-ray band as a
function of the multipole $\ell$: it is obtained after normalizing the
pixel values to their average of 4.06 $\times 10^{-12}$ \sbunits. The
solid line represents the average of the power spectra of the 10 maps;
the shaded region shows the corresponding r.m.s.  }
\label{fig:ps_soft}
\end{figure}

We show in Fig. \ref{fig:ps_soft} the power spectrum of the soft 
(0.5--2 keV) X-ray surface brightness arising from the hot ionized 
plasma; this power spectrum has been obtained in the same way as 
described in Section \ref{sect:power_sp} for the tSZ effect, but after 
dividing the pixel values for their average value corresponding to 
$4.06 \times 10^{-12}$ \sbunits. The amplitude of the r.m.s. is a 
direct consequence of the large variance between different fields of 
the soft X-ray emission coming from the LSS already discussed in 
\cite{roncarelli2006a}.

Confronting the soft X-ray power spectrum with the one of the tSZ 
effect shown in Fig. \ref{fig:pow_sp} we can see that the former peaks 
at higher multipoles than the latter. This is due to the fact that 
the dependence on the square of the gas density of the bremsstrahlung 
emission makes it more sensitive to the inner parts of collapsed objects.

\begin{figure}
\includegraphics[width=0.45\textwidth]{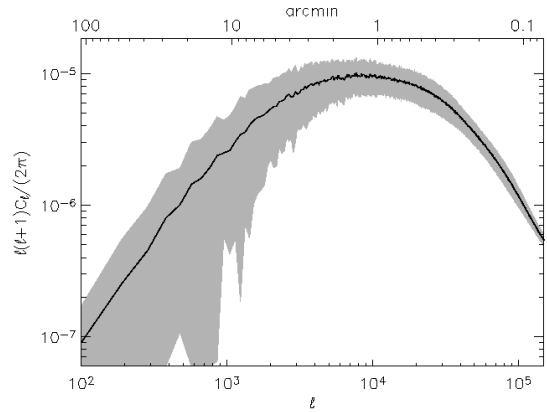}
\caption{
Power spectrum of the cross-correlation SXRB-tSZ 
at $\nu=30$ GHz as a function of the multipole $\ell$. The solid 
line represents the average of the power spectra of the 10 maps; 
the shaded region shows the corresponding r.m.s.
}
\label{fig:cps_soft-y}
\end{figure}

We show in Fig. \ref{fig:cps_soft-y} the power spectrum of the 
correlation of the tSZ signal with the soft X-ray emission; again, we 
computed them following the same methods described in Section
\ref{sect:power_sp} (using the set of 40 submaps for the tSZ-kSZ 
correlation) and assuming the frequency\footnote{For these
computations we dropped the negative sign arising from 
$g_\nu(x)\simeq-1.94$ for $\nu=30$ GHz.} $\nu=30$ GHz. As expected 
the cross-correlation 
peaks at an intermediate scale between the two power spectra, and it remains 
almost constant in the range between $\sim$0.5 and $\sim$3 arcmin. Therefore 
this constitutes the best angular range to study the two signals together.

Finally, we also compute the cross-correlation coefficient between the 
SXRB and the two SZ effects and show our results in Fig. 
\ref{fig:rl_Soft-yb} (left and right panel, respectively). The 
correlation with the tSZ signal is as high as $\sim$0.9 at 
$\ell \sim$1000,
slightly decreasing at lower angular scales. This result is in
contrast with that obtained by \cite{cheng2004} that predict a
correlation coefficient of about 0.3. The explanation can be found in
the different methods used to obtain the results: in fact
\cite{cheng2004} followed an analytical approach using several 
approximations to account for the cooling of the gas and 
for the shape of the density profiles. The correlation between the 
SXRB and the kSZ effect is also very high: in all of our light-cone 
realizations 
the values of $r_\ell$ for $\ell >$1000 are around 0.8 with negligible 
differences between them.


\section{Conclusions}\label{sect:conclu}

In this work we have studied the global properties of the tSZ and kSZ
effects using the results of a cosmological hydrodynamical simulation
of the \lcdm\ model. The simulation \citep{borgani2004} follows the
evolution of dark matter and baryons accounting for several physical
processes that affect the thermodynamical history of the gas: a time
dependent photo-ionizing uniform UV background, radiative cooling
processes and star formation with consequent feedback processes by
SN-II and galactic winds. We used the outputs of the simulations to
construct 10 independent light-cone realizations from $z=0$ out to
$z=6$ and computed associated mock maps, of size (3.78$^\circ$)$^2$,
of the Compton \ypar\ and Doppler \bpar. Using this dataset we
estimated the expected statistical properties of the two SZ effects,
we calculated their power spectra and studied their
cross-correlations. We also analised their correlation properties with
the soft (0.5--2 keV) X-ray emission obtained with the same light-cone
realizations and already studied in a previous work by
\cite{roncarelli2006a}.

Our main results can be summarized as follows.

\begin{enumerate}

\item The mean intensity of the Compton \ypar\ due to the IGM is 
      $<y>=(1.19 \pm 0.32) \times 10^{-6}$: almost 60 per 
      cent of this signal comes from WHIM and about half comes from 
      $z>1$. The distribution of the pixel values in 
      the sky is close to a lognormal, with variance increasing 
      at higher resolution up to\footnote{In a logarithmic scale.} 
      0.3 at the scales of 0.1 arcmin.

\item The peaks of the Doppler \bpar\, associated with collapsed 
      objects, are of the order of $|b|\simeq 10^{-5}$. This signal 
      presents a nearly gaussian distribution peaked around 0, with 
      a variance that increases with resolution out to the scales of 
      some arcsecond. The kSZ effect has a significant contribution 
      from high-redshift gas.

\begin{figure*}
\includegraphics[width=0.45\textwidth]{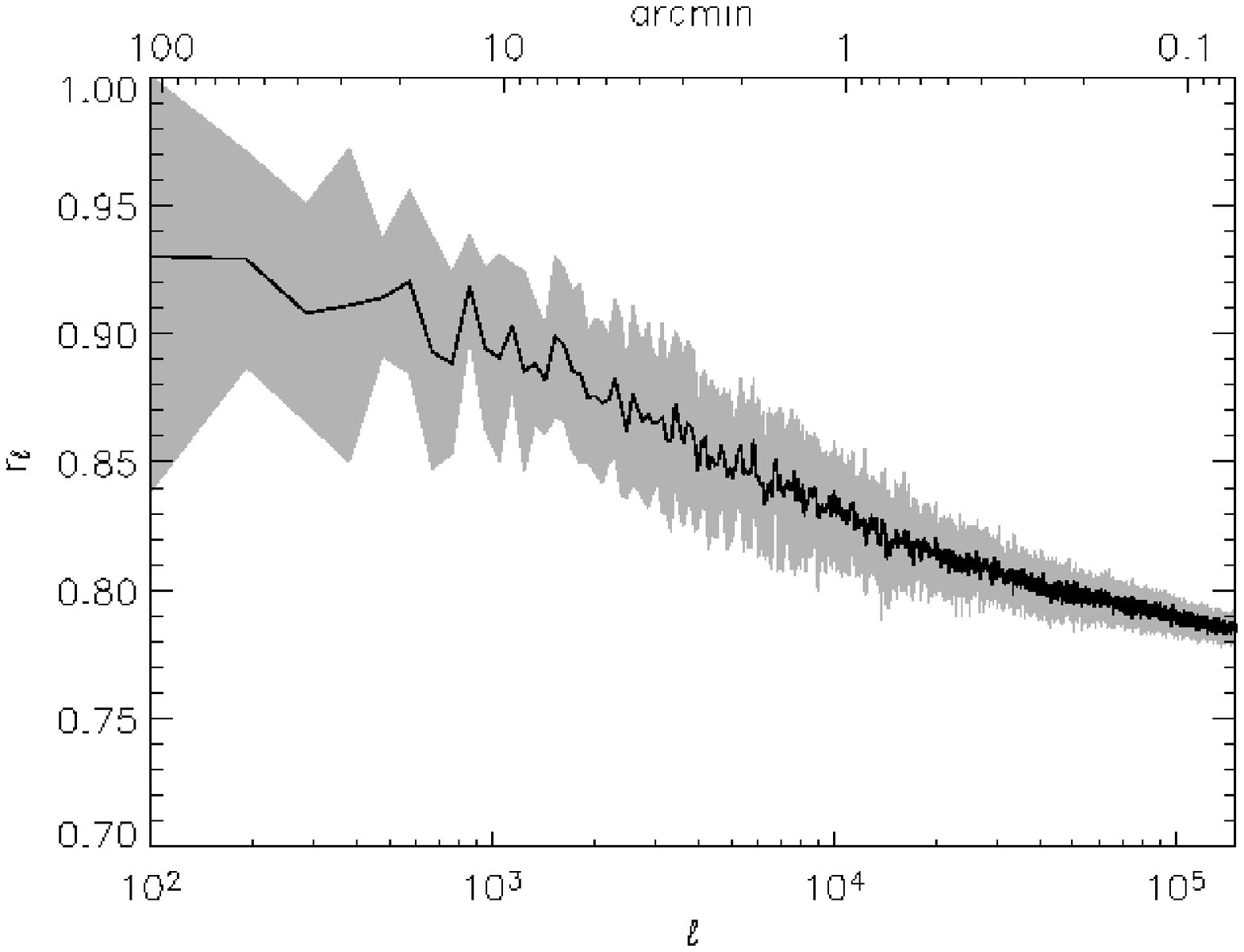}
\includegraphics[width=0.45\textwidth]{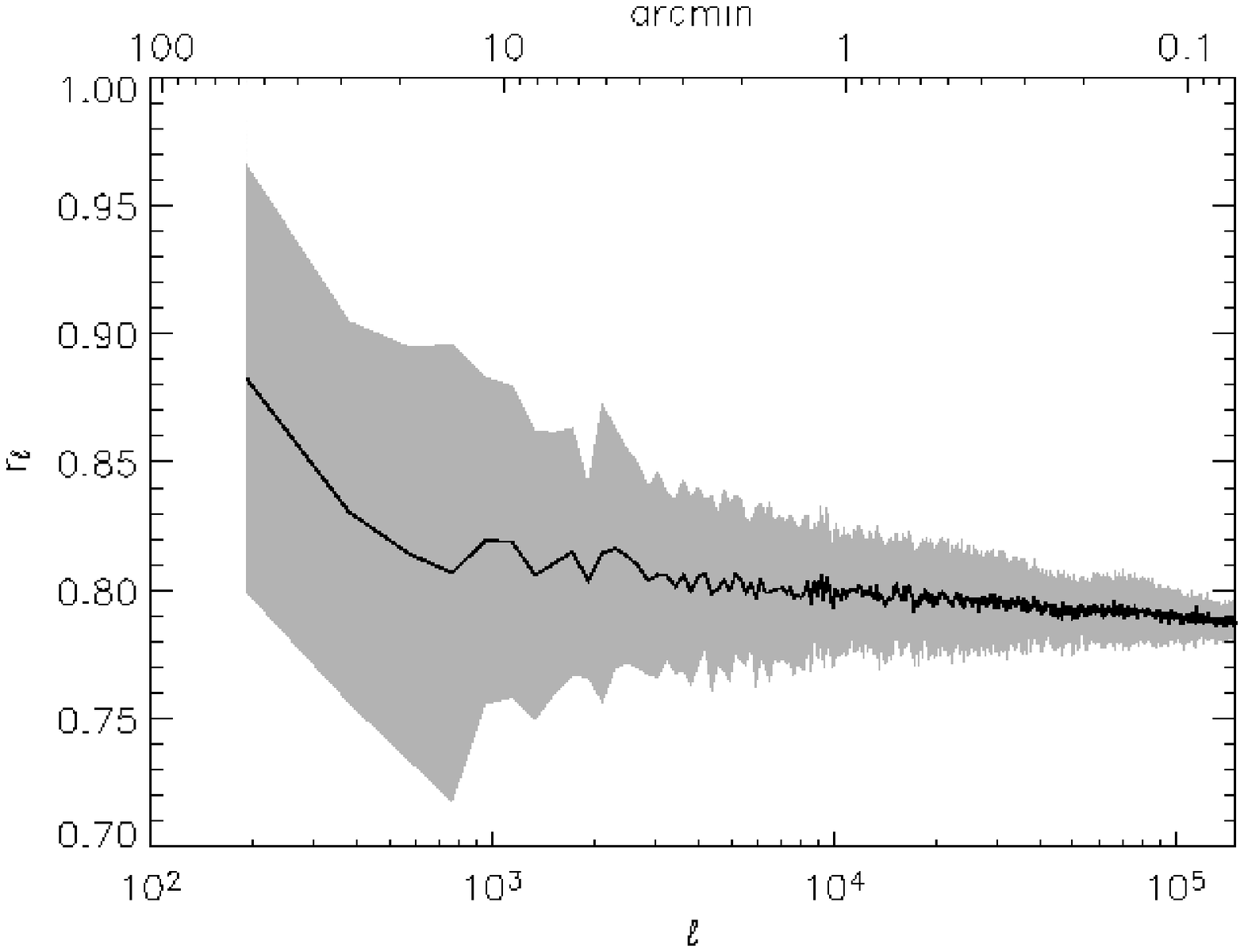}
\caption{
Cross-correlation coefficient between the SXRB in the (0.5--2 keV) band and 
the tSZ effect (left panel) and between the SXRB and kSZ effect 
(right panel) as a function of the multipole $\ell$. The solid line 
represents the average of the 10 (40) maps; the shaded region shows the 
r.m.s. between them.
}
\label{fig:rl_Soft-yb}
\end{figure*}

\item The power spectrum analysis shows that the tSZ effect dominates 
      the 
      primary CMB anisotropies for $\ell > 3000$ and peaks at $\ell 
      \simeq 5000$. The kSZ signal peaks at $\sim 1$ arcmin and has a flat 
      power spectrum towards high $\ell$ out to $\ell > 2 
      \times 10^5$. It dominates the tSZ signal for $\ell \gtrsim 
      7 \times 10^4$ at all frequencies. The two SZ effect are highly 
      correlated at all scales.

\item The SXRB is highly correlated with the two SZ effects, 
      particularly with the tSZ one ($r_\ell \simeq 0.8-0.9$). We 
      also 
      calculated a maximum-likelihood analytic formula that 
      connects the values of the two observables.

\end{enumerate}

In conclusion, our work further demonstrates the importance of the
upcoming measurements of both SZ effects and their complementarity 
with present-day and future X-ray data. A joint analysis of these
signals, in combination with high-resolution hydrodynamical
simulations, will allow one to obtain insights on the properties of
the baryonic component, and on the physical processes acting on it.

\section*{acknowledgements}

Computations have been performed by using the IBM-SP4/5 at CINECA 
(Consorzio Interuniversitario del Nord-Est per il Calcolo 
Automatico), Bologna, with CPU time assigned under an INAF-CINECA 
grant. We acknowledge financial contribution from contract ASI-INAF 
I/023/05/0. This work has been also partially supported by the PD-51 INFN 
grant. We wish to thank the anonymous referee for useful comments that 
improved the presentations of our
results. We acknowledge useful discussions with C. Baccigalupi, J. G. Bartlett, 
F. K. Hansen, A. Morandi, G. Murante and M. Righi. We also thank M. Bracchi 
for drawing Fig. \ref{fig:cubes}.

\newcommand{\noopsort}[1]{}

\label{lastpage}
\end{document}